\documentclass[preprint,preprintnumbers,amsmath,amssymb,floatfix,nofootinbib]
{revtex4}
\usepackage{graphicx}
\usepackage{url}
\begin{document}
\title{\boldmath{One-loop Corrections to the $S$ Parameter in the 
                 Four-site Model}}
\author{Sally Dawson$^{a}$}
\email[]{dawson@bnl.gov}
\author{C.~B.~Jackson$^{b}$}
\email[]{jackson@hep.anl.gov}
\affiliation{$^a$Department of Physics, Brookhaven National Laboratory, 
Upton, NY~ 11973, USA\\
$^b$ HEP Division, Argonne National Laboratory,9700 Cass Ave. Argonne, IL~60439 
\vspace*{.5in}}

\date{\today}

\begin{abstract}
We compute the leading chiral-logarithmic corrections to the $S$ parameter in
the four-site Higgsless model.  In addition to the usual electroweak gauge 
bosons of the Standard Model, this model contains two sets of 
heavy charged and neutral
gauge bosons.  In the continuum limit, the latter gauge bosons can be identified
with the first excited Kaluza-Klein states of the $W^\pm$ and $Z$ bosons
of a warped extra-dimensional model with an $SU(2)_L \times SU(2)_R \times
U(1)_X$ bulk gauge symmetry.  We consider delocalized fermions and show that
the delocalization parameter must be considerably tuned from its tree-level
{\it{ideal}} value in order to reconcile experimental constraints with the
one-loop results. Hence, the delocalization of fermions does not
solve the problem of large contributions to the $S$ parameter in this
class of theories and significant contributions to $S$ can potentially
occur at one-loop.  
\end{abstract}

\maketitle

\section{Introduction}
\label{sec:intro}

As the world awaits the turn-on of the Large Hadron Collider (LHC) at CERN, 
theorists and experimentalists alike are left to ponder the question of the source
of electroweak symmetry breaking (EWSB).  In the Standard Model (SM), the 
electroweak symmetry is spontaneously broken by a single $SU(2)$ scalar doublet
which acquires a non-zero vacuum expectation value (vev) and, subsequently, 
gives masses to the SM gauge bosons.  However, once radiative corrections are
included, the physical Higgs boson mass is found to be quadratically divergent
and extreme fine-tuning is required in order to achieve a mass on the order of 
hundreds of GeV.  This has become known as the {\it{large hierarchy problem}}.
Supersymmetric extensions of the SM can reduce the magnitude
of fine-tuning due to the presence of new particles in the loops contributing 
to the Higgs boson mass.

Alternatively, warped extra-dimensional (or Randall-Sundrum (RS)) models propose
to solve the large hierarchy problem by embedding the SM in an extra-dimensional
setup \cite{Randall:1999ee}.  In these models, the electroweak scale is generated 
from a large scale
(i.e., the Planck scale) through an exponential hierarchy.  The original version
of the RS model contained a slice of $AdS_5$ space bounded by two boundaries 
(or branes) where the SM was assumed to live on one of the boundaries.  Motivated
by the AdS/CFT correspondence \cite{Maldacena:1997re,Gubser:1998bc,
Witten:1998qj}, more recent versions of the RS scenario consider matter fields
and fermions of the SM propagating in the bulk, while the Higgs is constrained
to live on (or very near) the IR brane.  In order to avoid large tree-level
corrections to the $\rho$ parameter in these scenarios,
one must extend the bulk gauge group to a left-right symmetric form 
($SU(2)_L \times SU(2)_R \times U(1)$) \cite{Agashe:2003zs}.  Finally, it has 
been shown that the source of electroweak symmetry breaking in these models 
need not come from a fundamental scalar field.  In fact, by imposing certain 
boundary conditions on the gauge fields, one can give masses to the SM gauge 
bosons.  These models have been dubbed {\it{Higgsless models}} 
\cite{Csaki:2003dt,Cacciapaglia:2004rb,Nomura:2003du,Csaki:2003zu}.

In the gauge sector of these models, one expects a massless photon, light 
SM-like gauge bosons ($W^\pm$ and $Z^0$) plus towers of Kaluza-Klein (KK) 
partners to the light SM-like gauge bosons.  Thus, a strong constraint on 
these models comes from considering the $S$ parameter (since $S$ effectively
``counts'' the degrees of freedom in the electroweak 
sector)\cite{Altarelli:1990zd,Peskin:1991sw}.  In fact, the 
tree-level contributions to $S$ in these models can be quite large providing 
strong constraints on the KK gauge boson masses.  However, by judiciously 
choosing the localization of the light fermions in the bulk, the large 
tree-level corrections to $S$ can be cancelled \cite{Gherghetta:2000qt,
Huber:2000fh, Huber:2001gw,Davoudiasl:2000wi, Carena:2003fx,Chivukula:2005bn,Chivukula:2005xm,Foadi:2004ps}.  In light of 
this, it becomes imperative to assess higher-order corrections to $S$ in 
these models.

As in the SM, the one-loop corrections to gauge-boson self-energies in 
these models can be split into several gauge-invariant (and $R_\xi$ 
gauge-independent) pieces.  In other words, one can consider
the effects of new fermions, the Higgs boson (or other scalars) and 
the KK gauge bosons separately.  The effects of new fermions on the 
$S$ parameter in these models have been studied in 
Refs.~\cite{Carena:2004zn,Carena:2006bn,Carena:2007ua}, while the 
corrections from the Higgs sector have recently
been calculated in Ref.~\cite{Burdman:2008gm} using a ``holographic'' 
approach.  However, the effect of one-loop corrections to $S$ from
loops of gauge bosons in a model with a bulk gauge symmetry
$SU(2)_L \times SU(2)_R \times U(1)$ remains unknown.

The corrections to gauge-boson self-energies from loops of gauge bosons 
suffer from the fact that the final answer depends on the particular 
$R_\xi$ gauge that one uses to define the propagators of the gauge bosons
circulating in the loop.  This is a well-known problem that appears even 
in the SM \cite{Degrassi:1992ff}.  The remedy for this situation is to 
extract other gauge-dependent terms from vertex and box corrections and 
combine these with the corrections
to the two-point functions such that all $R_\xi$ gauge-dependent terms are
cancelled.  This process is known as the {\it{pinch technique}}
\cite{Cornwall:1981ru,Cornwall:1981zr,Cornwall:1989gv,Papavassiliou:1989zd}.  
It quickly becomes apparent that, in models with extra gauge
bosons in addition to the SM ones, the computation of the $S$ parameter
can be quite complicated.  Recently, however, a systematic algorithm for 
computing the one-loop corrections to $S$ from extra gauge bosons has been
developed in Ref.~\cite{Dawson:2007yk}.

In this paper, we consider the effects of KK gauge bosons on 
the $S$ parameter by studying the analogous effects in the deconstructed
four-site model \cite{Accomando:2008jh,SekharChivukula:2008gz}.
These models are generalizations of the BESS model with two new
triplets of gauge bosons\cite{Casalbuoni:1985kq,Casalbuoni:1986vq}.
The one-loop corrections to $S$ in the three-site model \cite{Perelstein:2004sc,
Foadi:2003xa,Chivukula:2006cg} have recently
been computed in Refs.~\cite{Matsuzaki:2006wn,Dawson:2007yk,
SekharChivukula:2007ic} and a fit to the electroweak data has been
performed in Ref.~\cite{Abe:2008hb}.  The four-site model is 
based on a linear moose diagram with a gauge structure of
$[SU(2)]^3 \times U(1)$.  The gauge groups are linked together via 
non-linear sigma model fields $\Sigma_i$ with link constants $f_i$.
Once the electroweak symmetry is broken, the gauge sector of this model
consists of a massless photon, light SM-like gauge bosons ($W^\pm$ and $Z^0$)
plus two sets of heavier gauge bosons ($\rho^\pm_i$ and $\rho^0_i$ with 
$i = 1,2$).  In the continuum limit, the latter can be thought of as the
first excited KK states of the SM $W^\pm$ and $Z^0$.  Additionally,
by choosing different values for the various link constants ($f_i$), 
one can mimic the warped nature of the extra dimension in RS-type models.

The paper is structured in the following way.  In Section \ref{sec:the-model},
we describe the model and calculate the mass eigenvalues and eigenvectors
for both the charged and neutral gauge bosons. In Section III, we present
the chiral logarithmic corections to the gauge boson self energies from 
gauge boson self interactions and the resulting contribution to the $S$ 
parameter is given in Section IV.  We conclude with some observations in
Section V. 
 
\section{The Model}
\label{sec:the-model}

\begin{figure}[t]
\begin{center}
\includegraphics[scale=1.0]{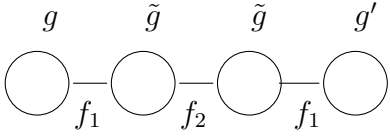}
\end{center}
\caption[]{The moose diagram for the four-site model.}
\label{fg:4SMmoose}
\end{figure}

In this section, we describe in detail the four-site model \cite{Accomando:2008jh,
SekharChivukula:2008gz}.  The model is based on an $SU(2)_L \times SU(2)_{V_1} \times
SU(2)_{V_2} \times U(1)$ gauge symmetry and is depicted by the moose diagram shown in
Fig.~\ref{fg:4SMmoose}. The corresponding gauge fields are $L^\mu$,
$V_1^\mu$, $V_2^\mu$, and $R^\mu$, with gauge couplings $g, \tilde{g},
{\tilde g}$ and $g^\prime$ respectively. 

The Lagrangian for the model consists of several parts.  First, the non-linear 
sigma model terms are given by:
\begin{equation}
{\cal{L}}_{n\ell\sigma m} = \sum_{i=1}^3 \frac{f_i^2}{4} \biggl[
  \mbox{Tr} D^\mu \Sigma_i D_\mu \Sigma_i^\dagger \biggr]\,,
\label{eq:Lnlsm}
\end{equation}
where the covariant derivatives are defined as:
\begin{eqnarray}
\label{eq:DmuSig1}
D_\mu \Sigma_1 &=& \partial_\mu \Sigma_1 - i g L_\mu \Sigma_1 + 
                   i \tilde{g} \Sigma_1 V_{1 \mu} \\
\label{eq:DmuSig2}
D_\mu \Sigma_2 &=& \partial_\mu \Sigma_2 - i \tilde{g} V_{1 \mu} \Sigma_2 + 
                   i \tilde{g} \Sigma_2 V_{2 \mu} \\
\label{eq:DmuSig3}
D_\mu \Sigma_3 &=& \partial_\mu \Sigma_3 - i \tilde{g} V_{2 \mu} \Sigma_3 + 
                   i g^\prime \Sigma_3 R_{\mu} \,,
\end{eqnarray}
with:
\begin{equation}
L_\mu = T^a L^a_\mu \,\,\, ; \,\,\, V_{i \mu} = T^a V_{i \mu}^a
         \,\,\, ; \,\,\, R_\mu = T^3 B_\mu \,,
\end{equation}
and $T^a = \frac{\sigma^a}{2}$, where $\sigma^a$ are the usual Pauli matrices.
Note that, in general, the link constants $f_i$ are free to take any value.  In this
paper, however, we will consider two distinct possibilities: one where all $f_i$ are 
equal and one where the middle link constant takes a different value from the other 
two (which are set equal to each other).  In the continuum limit, the former corresponds
to a flat extra-dimension, while the latter corresponds to a warped extra-dimension.  

The gauge-boson kinetic terms are given by:
\begin{equation}
{\cal{L}}_{g} = -\frac{1}{2}\, \mbox{Tr} [L^{\mu\nu}]^2 - 
   \frac{1}{2}\sum_{i=1}^{2}\, \mbox{Tr} [V_i^{\mu\nu}]^2 - 
   \frac{1}{2} \, \mbox{Tr} [R^{\mu\nu}]^2 \,
\label{eq:Lgauge}
\end{equation}
where $L_{\mu\nu}$, $V_i^{\mu\nu}$ and $R^{\mu\nu}$ are the matrix field-strengths
of the four gauge groups.

Next, we consider the couplings of light fermions to the various gauge groups.  With
the fermions completely localized to the two end sites, the four-site model generates 
a large correction to the $S$ parameter at tree-level.  However, if one allows the fermions 
to have small, non-zero couplings to the interior sites, this large tree-level 
contribution to $S$ can be cancelled completely.  It has been shown in 
Refs.~\cite{Chivukula:2005bn,Accomando:2008jh} that, in general, it is enough to consider one-site
delocalization in order to cancel the large tree-level contributions.  Thus, we 
assume that the light fermions couple mainly to the two end groups, as well as 
a small coupling to $SU(2)_{V_1}$ such that the Lagrangian takes the form:
\begin{equation}
{\cal{L}}_f = -g^\prime \bar{\psi} \gamma_\mu (Y_L P_L + Y_R P_R) B^\mu \psi - 
  g (1 - x_1) \bar{\psi} \gamma_\mu T^a L^{a,\mu} P_L \psi -
  \tilde{g} x_1 \bar{\psi} \gamma_\mu T^a V_1^{a,\mu} P_L \psi \,,
\label{eq:Lfermion}
\end{equation}
where $P_{L,R}= \frac{1}{2}(1\mp \gamma_5)$, the
electromagnetic charge is related to the isospin, $Q_{em}=T_3+Y$ and 
the parameter $x_1$ measures the amount of delocalization and is assumed
to be $0 < x_1 \ll 1$.  We note that this expression is not separately 
gauge-invariant under $SU(2)_L$ and $SU(2)_{V_1}$.  Rather, the fermions should
be viewed as being charged under $SU(2)_L$ and the terms proportional $x_1$ 
arise from an operator of the form:
\begin{equation}
{\cal{L}}_f^\prime = - x_1 \, {\overline{\psi}}
 \gamma^\mu (i D_\mu \Sigma_1 \Sigma^\dagger_1)
  P_L \psi \,.
\end{equation}

In addition to the terms listed above, one must also include higher-derivative operators
since the theory is non-renormalizeable.  In particular, several ${\cal O}(p^4)$
 operators
that one can write down are relevant to the $S$ parameter.  Expressing these in terms of
the four-site model gauge fields, the relevant operators are \cite{Perelstein:2004sc}:
\begin{equation}
{\cal{L}}_4 = 
  c_1 \tilde{g}g^\prime \, \mbox{Tr} [V_{2,\mu\nu} \Sigma_3 B^{\mu\nu} T^3 \Sigma_3^\dagger] +
  c_2 \tilde{g}g \, \mbox{Tr} [V_{1,\mu\nu} \Sigma_1^\dagger L^{\mu\nu} \Sigma_1] +
  c_3 \tilde{g}^2 \, \mbox{Tr} [V_{2,\mu\nu} \Sigma_2^\dagger V_1^{\mu\nu} \Sigma_2] \,.
\label{eq:L4doperators}
\end{equation}

Finally, Ref.~\cite{SekharChivukula:2008gz} has shown that by including an $L_{10}$-like
mixing between the middle two sites:
\begin{equation}
{\cal{L}}_\epsilon = -\frac{\epsilon}{2} \, \mbox{Tr} [V_{1,\mu\nu} \Sigma_2 V_{2,\mu\nu}
  \Sigma_2^\dagger] \,,
\label{eq:Lepsilon}
\end{equation}
one can cancel the dangerously large tree-level contributions to the $S$ parameter 
without delocalizing the light fermions.  To avoid ghosts, one must require that the
free parameter $\epsilon$ satisfy $|\epsilon| < 1$.  However, one finds that the 
value of $\epsilon$ required to cancel the $S$ parameter at tree-level is of
order one which is much larger than would be expected from naive dimensional analysis
\cite{Georgi:1992dw,Agashe:2007mc}.  In the following, therefore, 
we will neglect this term and 
study the simpler version of the four site model
 given by the sum of Eqs.~(\ref{eq:Lnlsm}), 
(\ref{eq:Lgauge}), (\ref{eq:Lfermion}) and (\ref{eq:L4doperators}).

The model approximates the SM in the limit:
\begin{equation}
x \equiv \frac{g}{\tilde{g}} \ll 1, \,\,\, y \equiv \frac{g^\prime}{\tilde{g}} \ll 1 \,,
\label{eq.xydefs}
\end{equation}
in which case we expect the spectrum in the gauge sector to consist of a massless
photon, light SM-like $W$ and $Z$ bosons and  two sets of heavy bosons which we denote as
$\rho^\pm_i$ and $\rho^0_i$ with $i = 1,2$.  The four-site model couplings $g$ and $g^\prime$ 
are then numerically equal to the SM $SU(2)_L$ and $U(1)_Y$ gauge couplings respectively
in this limit.  We therefore define an angle $\theta$ such that:
\begin{equation}
g^2 \simeq \frac{4\pi\alpha}{s^2} = \frac{e^2}{s^2}, \,\,\,
g^{\prime 2} \simeq \frac{4\pi\alpha}{c^2}, \,\,\,
\frac{s}{c} = \frac{g^\prime}{g}\,,
\label{eq:SMcoups}
\end{equation}
where $s(c) = \sin\theta (\cos\theta)$, $\alpha$ is the fine-structure constant and
$e$ is the charge of the electron.
 
\subsection{Mass Eigenstates and Their Interactions}
\label{subsec:mass-estates}

In unitary gauge (where $\Sigma_i \equiv {\cal{I}}$), the quadratic piece of the full
Lagrangian gives rise to mass terms for the neutral
($M_{NC}$) and charged ($M_{CC}$) gauge bosons of the form:

\begin{equation}
{\cal{L}} = \frac{1}{2}\sum_{i=0}^3 W_{i,\mu} M_{ij,NC}^2 W^\mu_j
+\sum_{i=0}^2 W_{i,\mu} M_{ij,CC}^2 W^\mu_j\,,
\label{eq:Lmass}
\end{equation}
where, in ``site'' space, the vectors $W_{i,\mu}$ are given by:
\begin{eqnarray}
\left(
\begin{array}{c}
L_{\mu} \\ 
V_{1,\mu}  \\
V_{2,\mu} \\
B_{\mu}
\end{array}
\right) \,.
\label{eq:Wvector}
\end{eqnarray}
The eigenstates corresponding to Eq.~(\ref{eq:Lmass})
satisfy the eigenvalue equation:
\begin{equation}
M^2 \vec{v}_n = m_n^2 \vec{v}_n \,,
\label{eq:EigenVal}
\end{equation}
where $\vec{v}_n$ is a vector in site space with components $v_n^i$.  The superscript $i$ 
labels the sites, running from 0 to 2 for the charged bosons ($n = W^\pm , \rho_1^\pm , 
\rho_2^\pm$), and 0 to 3 for neutral ones ($n = A, Z^0, \rho_1^0 , \rho_2^0$).  Then, choosing
eigenvectors normalized by $\vec{v}^T_n \vec{v}_m = \delta_{mn}$, the gauge
eigenstates ($W_\mu^i$) and mass eigenstates ($W^\prime_{n\mu}$) are related by:
\begin{equation}
W_\mu^i = \sum_n v_n^i W^\prime_{n\mu} \,.
\label{eq:Eigenvecs}
\end{equation}

\subsubsection{The Charged Sector}
\label{subsubsec:charged-mass-estates}

First, we consider the charged gauge boson sector.  The mass matrix in this sector 
takes the form: 
\begin{eqnarray}
{\cal{M}}_{CC}^2 & = & 
\frac{\tilde{g}^2}{4}
\left(
\begin{array}{ccc} 
x^2 f_1^2 & -x f_1^2 & 0 \\
-x f_1^2 & f_1^2 + f_2^2 & -f_2^2\\
0 & -f_2^2 & f_1^2 + f_2^2 
\end{array}
\right) \,.
\end{eqnarray}
Since we are interested in the limit $x = g/\tilde{g} \ll 1$, we diagonalize the mass
matrices perturbatively in $x$.  To ${\cal{O}}(x^2)$, we find the mass of the SM-like
$W$ boson is:
\begin{equation}
M_W^2 \simeq \frac{g^2}{4} \frac{f_1^2 f_2^2}{f_1^2 + 2 f_2^2} 
  \biggl( 1 - x^2 z_W \biggr) \,,
\label{eq:Mwwarped}
\end{equation}
while the masses of the heavier charged gauge bosons are:
\begin{equation}
M_{\rho^\pm_1}^2 \simeq \frac{\tilde{g}^2 f_1^2}{4} 
  \biggl( 1 + \frac{x^2}{2} \biggr) \,,
\label{eq:Mrhopm1warped}
\end{equation}
and:
\begin{equation}
M_{\rho^\pm_2}^2 \simeq \frac{\tilde{g}^2 (f_1^2 + 2f_2^2)}{4} 
  \biggl( 1 + \frac{x^2}{2} z^4 \biggr) \,,
\label{eq:Mrhopm2warped}
\end{equation}
where:
\begin{eqnarray}
\label{eq:zdef}
z &=& \frac{f_1}{\sqrt{f_1^2 + 2 f_2^2}} \,, \\
\nonumber\\
\label{eq:zWdef}
z_W &=& \frac{f_1^4 + 2f_1^2 f_2^2 + 2 f_2^4}{(f_1^2 + 2 f_2^2)^2} = 
  \frac{1}{2}(1 + z^4) \,.
\end{eqnarray}
For future reference, we note that the ratios of the masses are given by (for small
values of $x^2$):
\begin{eqnarray}
\frac{M_W^2}{M_{\rho^\pm_1}^2} &\simeq& x^2 \left( \frac{1-z^2}{2} \right) \,,
\nonumber\\
\frac{M_W^2}{M_{\rho^\pm_2}^2} &\simeq& x^2 \left(\frac{z^2(1-z^2)}{2}\right) \,,
\nonumber\\
\frac{M_{\rho^\pm_1}^2}{M_{\rho^\pm_2}^2} &\simeq&
  z^2 \left(1 + \frac{x^2}{2}(1-z^4) \right)\,.
\end{eqnarray}

Finally, we expand the gauge-eigenstate fields in terms of the mass eigenstates
as:
\begin{eqnarray}
\label{eq:Lpmeigen}
L_\mu^\pm &=& v_{W^\pm}^L W^\pm_\mu + v_{\rho_1^\pm}^L \rho_{1,\mu}^\pm +
  v_{\rho_2^\pm}^L \rho_{2,\mu}^\pm \,,\\
\nonumber\\
\label{eq:V1pmeigen}
V_{1,\mu}^\pm &=& v_{W^\pm}^{V_1} W^\pm_\mu + v_{\rho_1^\pm}^{V_1} \rho_{1,\mu}^\pm +
  v_{\rho_2^\pm}^{V_1} \rho_{2,\mu}^\pm \,,\\
\nonumber\\
\label{eq:V2pmeigen}
V_{2,\mu}^\pm &=& v_{W^\pm}^{V_2} W^\pm_\mu + v_{\rho_1^\pm}^{V_2} \rho_{1,\mu}^\pm +
  v_{\rho_2^\pm}^{V_2} \rho_{2,\mu}^\pm \,.
\end{eqnarray}
We give the explicit expressions for the eigenvector components in 
Appendix~\ref{app:eigenvec-components}.

\subsubsection{The Neutral Sector}
\label{subsubsec:neutral-mass-estates}

The mass matrix for the neutral gauge fields takes the form:
\begin{eqnarray}
{\cal{M}}_{NC}^2 & = & 
\frac{\tilde{g}^2}{4}
\left(
\begin{array}{cccc} 
x^2 f_1^2 & -x f_1^2 & 0 & 0 \\
-x f_1^2 & f_1^2 + f_2^2 & -f_2^2 & 0\\
0 & -f_2^2 & f_1^2 + f_2^2 & -x t f_1^2 \\
0 & 0 & -x t f_1^2 & x^2 t^2 f_1^2  
\end{array}
\right) \,,
\end{eqnarray}
where $t \equiv \tan\theta$.  Again, we diagonalize the mass matrix perturbatively
in $x$.  We find one zero eigenvalue corresponding to the photon as well as three
massive states:
\begin{eqnarray}
\label{eq:Mzwarped}
M_Z^2 &\simeq& \frac{g^2}{4 c^2} \frac{f_1^2 f_2^2}{f_1^2 + 2 f_2^2}
  \biggl(1 - x^2 z_Z \biggr) \,, \\
\nonumber\\
\label{eq:Mrho01warped}
M_{\rho^0_1}^2 &\simeq& \frac{\tilde{g}^2 f_1^2}{4} 
  \biggl(1 + \frac{x^2}{2 c^2} \biggr) \,, \\
\nonumber\\
\label{eq:Mrho02warped}
M_{\rho^0_2}^2 &\simeq& \frac{\tilde{g}^2 (f_1^2 + 2f_2^2)}{4}
  \biggl(1 + \frac{x^2 z^4}{2 c^2} \biggr)\,, 
\end{eqnarray}
where:
\begin{equation}
z_Z = \frac{1}{2} \frac{(z^4 + \cos^2 2\theta)}{\cos^2 \theta} \,.
\label{eq:zZ}
\end{equation}

Next, we expand the gauge-eigenstate fields in terms of the mass eigenstates as:
\begin{eqnarray}
\label{eq:L3eigen}
L_\mu^3 &=& v_A^L A_\mu + v_Z^L Z_\mu + v_{\rho^0_1}^L \rho^0_{1,\mu} + 
  v_{\rho^0_2}^L \rho^0_{2,\mu} \,, \\
\label{eq:V13eigen}
V_{1,\mu}^3 &=& v_A^{V_1} A_\mu + v_Z^{V_1} Z_\mu + v_{\rho^0_1}^{V_1} \rho^0_{1,\mu} + 
  v_{\rho^0_2}^{V_1} \rho^0_{2,\mu} \,, \\
\label{eq:V23eigen}
V_{2,\mu}^3 &=& v_A^{V_2} A_\mu + v_Z^{V_2} Z_\mu + v_{\rho^0_1}^{V_2} \rho^0_{1,\mu} + 
  v_{\rho^0_2}^{V_2} \rho^0_{2,\mu} \,, \\
\label{eq:Beigen}
B_\mu &=& v_A^B A_\mu + v_Z^B Z_\mu + v_{\rho^0_1}^B \rho^0_{1,\mu} + 
  v_{\rho^0_2}^B \rho^0_{2,\mu} \,.
\end{eqnarray}
The expressions for the individual $v_i^j$'s are given in 
Appendix~\ref{app:eigenvec-components}.

\section{One-loop Corrections to the Gauge Boson Self-energies}
\label{sec:loop-corrections}

In this section, we compute the one-loop corrections needed to calculate the 
$S$ parameter in the four-site model.  The $S$ parameter is defined in 
the mass eigenstate basis as \cite{Peskin:1991sw}:
\begin{equation}
\frac{\alpha \Delta S}{4 s^2 c^2} = \Delta \Pi_{ZZ}(M_Z^2) - \Delta \Pi_{AA}(M_Z^2)
  - \frac{c^2 - s^2}{cs} \Delta \Pi_{ZA}(M_Z^2) \,,
\label{eq:Sdef}
\end{equation}
where:
\begin{equation}
\Delta \Pi_{ij}(M_Z^2) \equiv \frac{\Pi_{ij}(M_Z^2) - \Pi_{ij}(0)}{M_Z^2} \,,
\label{eq:DeltaPidef}
\end{equation}
and our convention for the self-energies is:
\begin{equation}
i\Pi_{ij}^{\mu\nu}(q^2) = g^{\mu\nu} \Pi_{ij}(q^2) + (q^\mu q^\nu \,\, \mbox{term})\,.
\end{equation}

Since the four-site model is a non-renormalizable theory, the one-loop corrections
will result in expressions which are divergent.  In other words, calculating the 
one-loop corrections using dimensional regularization (in $d = 4 - 2\epsilon$ dimensions)
the resulting expressions contain terms which diverge as $1/\epsilon$.  Alternatively,
if we were to compute the one-loop corrections using a momentum cutoff, the divergent
terms are logarithms of the form $\log \frac{\Lambda^2}{M^2}$ where $\Lambda$ is assumed
to be the cutoff scale of the effective theory and $M$ is the heaviest of the masses
circulating in the loop.  If the hierarchy between $\Lambda$ and $M$ is large, then the
contributions from these {\it{chiral logarithms}} dominates over any finite terms.  Below,
we will compute the leading chiral-logarithmic contributions to the $S$ parameter 
in the four-site model.

\subsection{The Two-point Functions}
\label{subsec:two-pt-corrections}

\begin{figure}[t]
\begin{center}
\includegraphics[bb=159 546 430 721,scale=0.5]{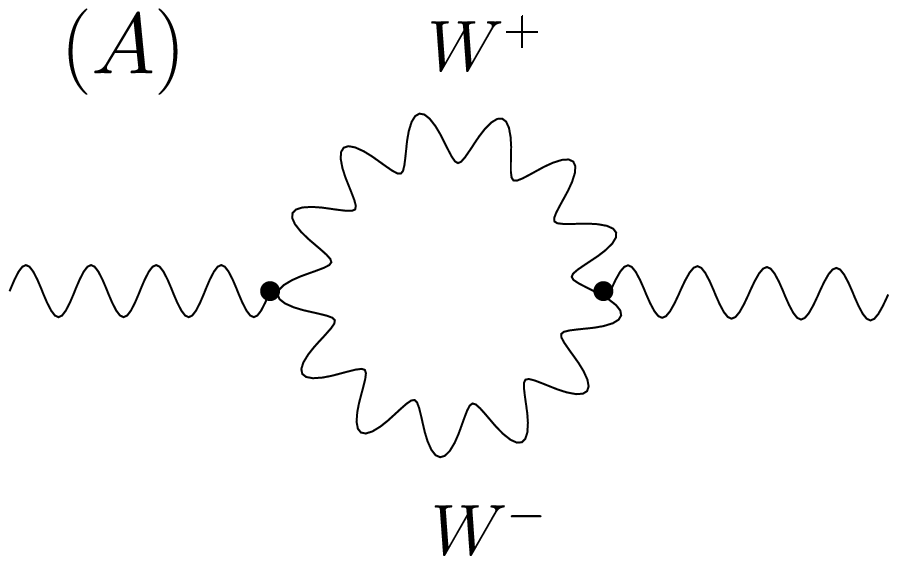} 
\includegraphics[bb=159 546 430 721,scale=0.5]{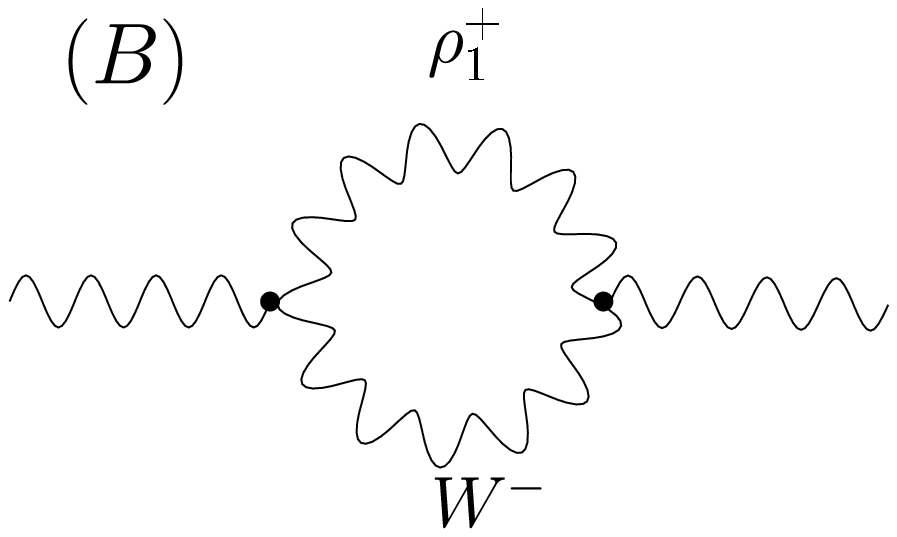}
\includegraphics[bb=159 546 430 721,scale=0.5]{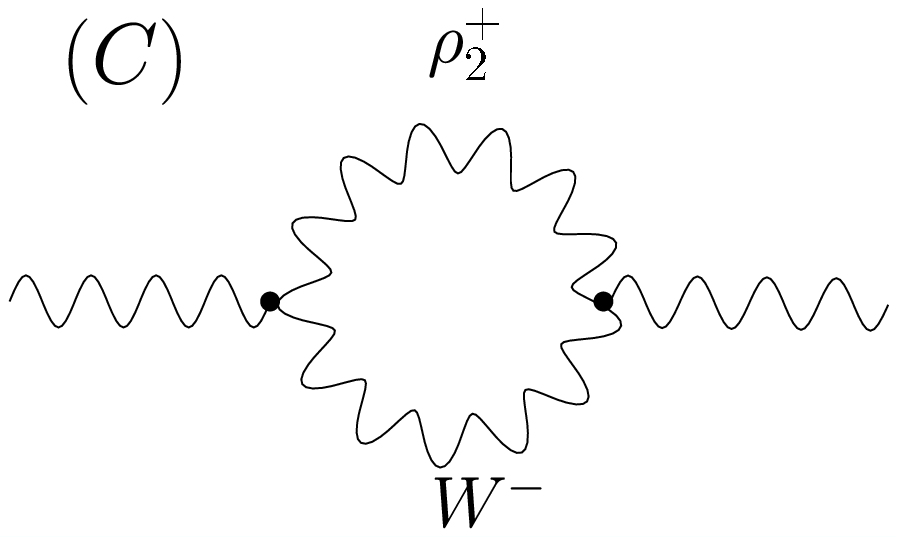} \\ 
\vspace{0.75cm}
\includegraphics[bb=159 546 430 721,scale=0.5]{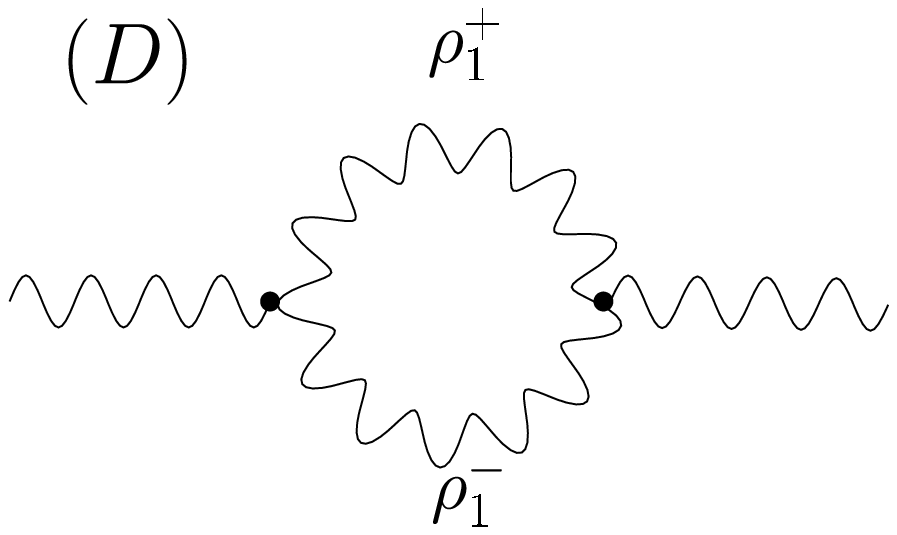}
\includegraphics[bb=159 546 430 721,scale=0.5]{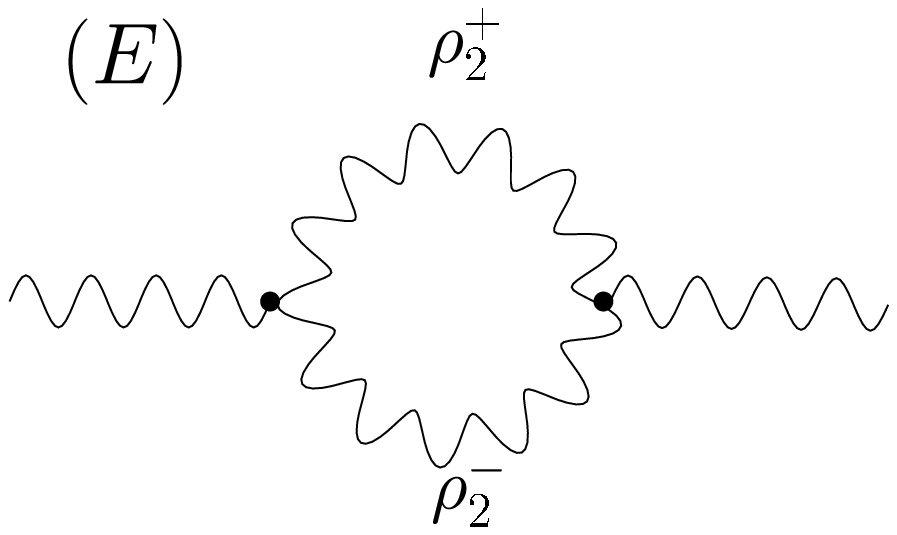}
\includegraphics[bb=159 546 430 721,scale=0.5]{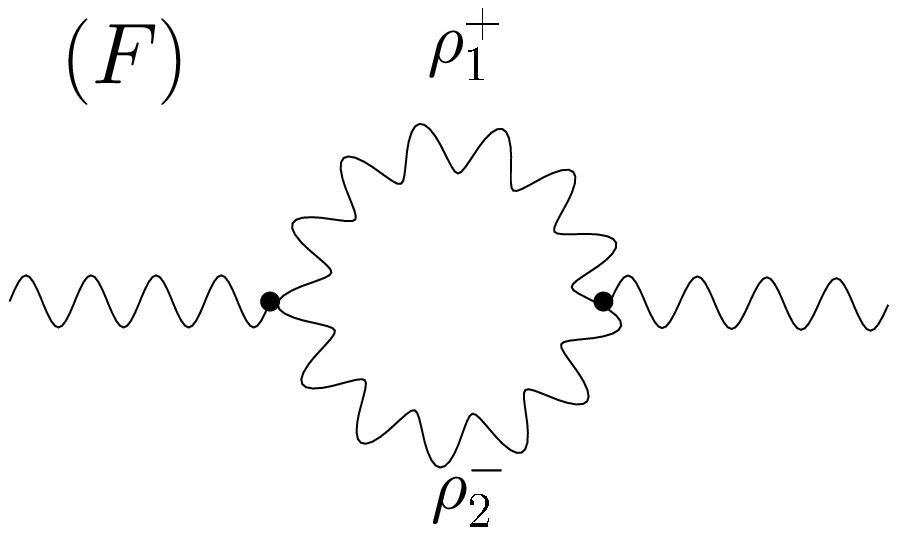}
\end{center}
\caption[]{One-loop corrections to the two-point functions $\Pi_{AA},
  \Pi_{ZZ}$ and $\Pi_{ZA}$ in the four-site model.}
\label{fg:twopt-diagrams}
\end{figure}

\begin{figure}[t]
\begin{center}
\includegraphics[bb=187 588 413 728,scale=0.6]{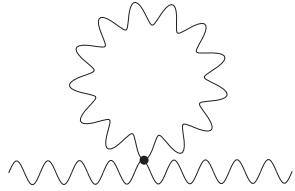}
\end{center}
\caption[]{Diagram containing four-point couplings between gauge bosons.}
\label{fg:4pt-diagram}
\end{figure}

The diagrams which contribute to the gauge-boson two-point functions in the four-site
model are shown in Fig.~\ref{fg:twopt-diagrams}.  Note that diagrams which contain
four-point interactions such as Fig.~\ref{fg:4pt-diagram} do not contain any 
$q^2$-dependence (where $q$ is the momentum of the external gauge bosons).  Hence, 
given Eq.~(\ref{eq:DeltaPidef}), these diagrams do not contribute to $S$ and we 
will neglect them in the following.  Note also that we are working in unitary gauge
where only physical particles contribute to the loops.

The individual contributions from the diagrams in Fig.~\ref{fg:twopt-diagrams} to the
gauge-boson two-point functions are summarized in Appendix~\ref{app:feyngraph-results}.
Summing the contributions to the photon two-point function, we find:
\begin{equation}
\label{eq:DelPiAA-twopt}
\Delta\Pi_{AA}^{two-pt.} = \frac{\alpha}{4\pi} \biggl[
  \biggl( 7 - \frac{7}{6 c^2} - \frac{1}{12 c^4} \biggr) \log \frac{\Lambda^2}{M_W^2} +
  7 \log \frac{\Lambda^2}{M_{\rho_1^\pm}^2} +
  7 \log \frac{\Lambda^2}{M_{\rho_2^\pm}^2}
  \biggr] \,.
\end{equation}
  
The total contribution to the $Z$-photon mixing amplitude $\Delta\Pi_{ZA}$ is:
\begin{equation}
\label{eq:DelPiZA-twopt}
\Delta\Pi_{ZA}^{two-pt.} = \frac{\alpha}{4\pi cs} \biggl[
  c^2 \biggl( 7 - \frac{7}{6 c^2} - \frac{1}{12 c^4} \biggr) \log \frac{\Lambda^2}{M_W^2} +
  \frac{7(c^2-s^2)}{2} \log \frac{\Lambda^2}{M_{\rho_1^\pm}^2} +
  \frac{7(c^2-s^2)}{2} \log \frac{\Lambda^2}{M_{\rho_2^\pm}^2}
  \biggr] \,.
\end{equation}

Finally, the total contribution to the $Z$ boson two-point function from the 
diagrams in Fig.~\ref{fg:twopt-diagrams} is:
\begin{eqnarray}
\Delta\Pi_{ZZ}^{two-pt.} &=& \frac{\alpha}{4\pi s^2 c^2} \biggl[
  c^4 \biggl(7 - \frac{7}{6c^2} - \frac{1}{12c^4} \biggr) 
  \log\frac{\Lambda^2}{M_W^2} \nonumber\\
&& \,\,\,\,\,\,\,\,\,\,\,\,\,\,\,
 + \left[
  \frac{17}{24}\frac{(1-z^4)^2}{(1-z^2)} + \frac{7(c^2-s^2)^2}{4}
   \right] \log \frac{\Lambda^2}{M_{\rho_1^\pm}^2} \nonumber\\
&& \,\,\,\,\,\,\,\,\,\,\,\,\,\,\,
  + \left[
  \frac{17}{24}z^2 (1+z^4) + \frac{25}{12}z^4 + \frac{7(c^2-s^2)^2}{4}
    \right] \log \frac{\Lambda^2}{M_{\rho_2^\pm}^2} \biggr]\,.
\label{eq:DelPiZZ-twopt}
\end{eqnarray}

\subsection{Pinch Contributions and the Total Self-energies}
\label{subsec:pinch-contributions}

As discussed earlier, the one-loop corrections to the two-point functions from 
loops of gauge bosons exhibit non-trivial dependence on the particular $R_\xi$
gauge used to define the gauge boson propagators.  This gauge-dependence carries
over into the calculation of observables such as the oblique parameters resulting
in gauge-dependent expressions for $S, T$ and $U$.  The remedy for this situation
is to isolate gauge-dependent terms from other one-loop corrections (i.e., vertex
and box corrections) and combine these with the one-loop corrections to the two-point
functions.  The result is a gauge-independent expression which can reliably be 
compared to experimental data.  This technique, known as the Pinch Technique (PT),
was first developed for the SM, but, recently, an algorithm has been developed to
extend the PT to theories with extra gauge bosons beyond those of the SM
\cite{Dawson:2007yk}.  This algorithm was utilized to compute $S$ and $T$ at one-loop
in the three-site model and shown to produce identical results to those of 
Refs.~\cite{Matsuzaki:2006wn,SekharChivukula:2007ic} which were obtained using
different methods.  In this section, we compute the ``pinch'' contributions in the 
four-site model.

\begin{figure}[t]
\begin{center}
\includegraphics[bb=179 544 436 727,scale=0.5]{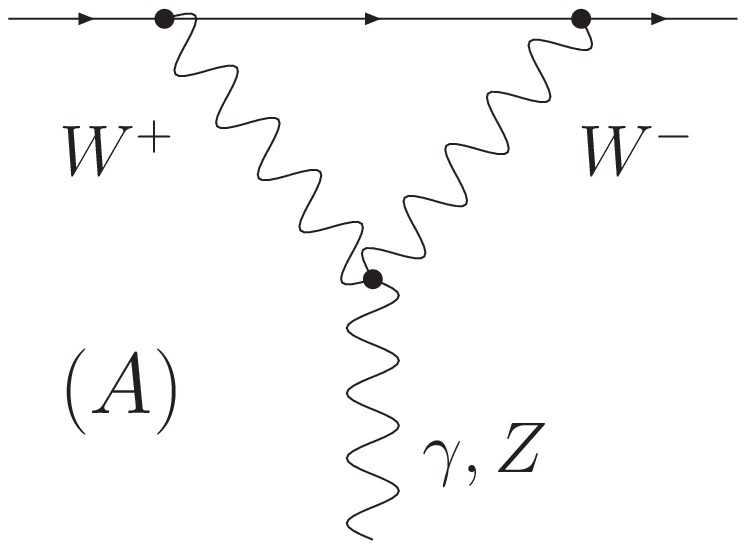} 
\includegraphics[bb=179 544 436 727,scale=0.5]{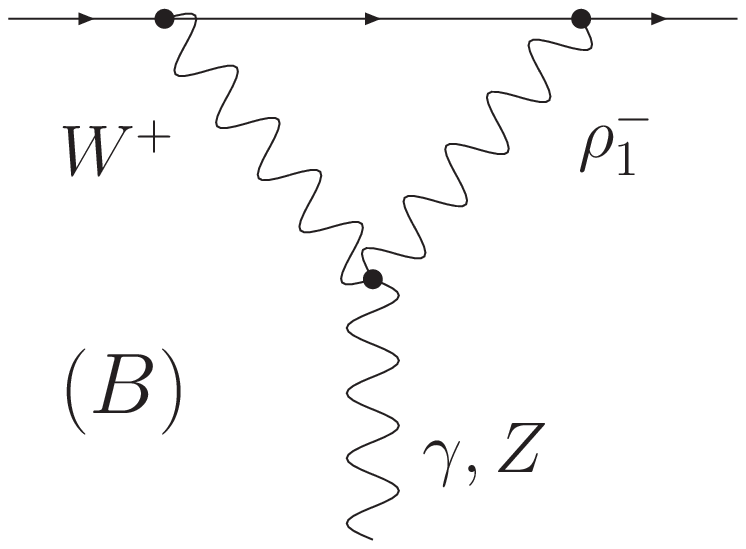}
\includegraphics[bb=179 544 436 727,scale=0.5]{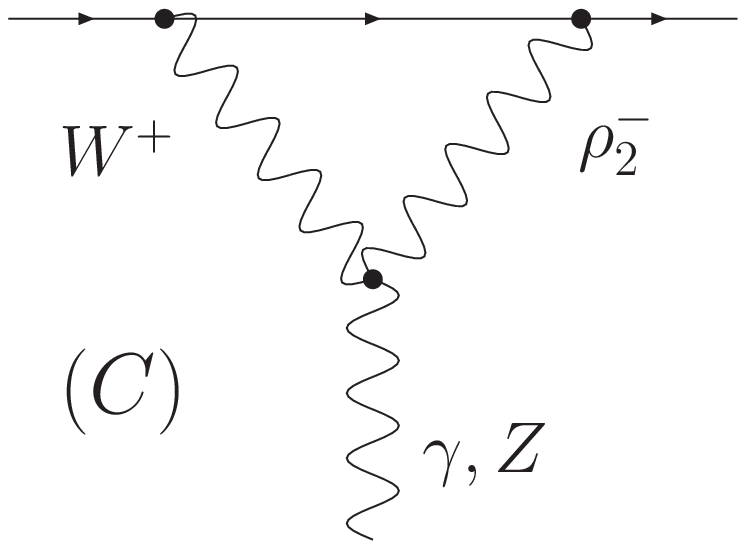} \\ 
\vspace{0.75cm}
\includegraphics[bb=179 544 436 727,scale=0.5]{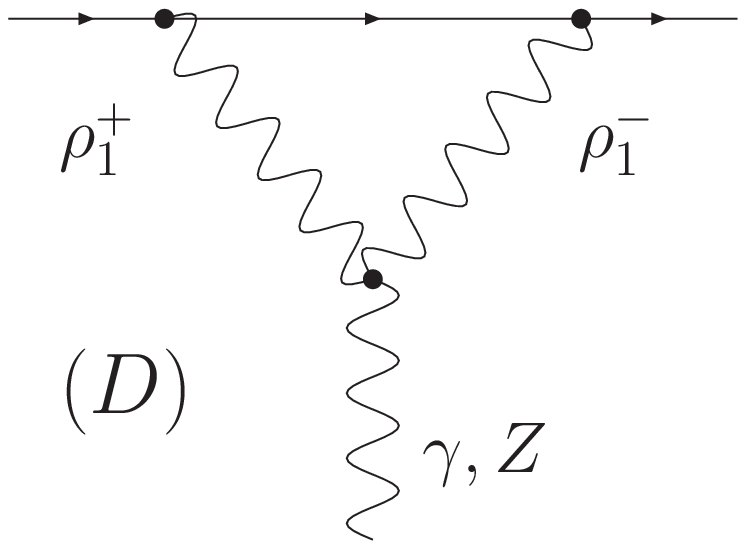}
\includegraphics[bb=179 544 436 727,scale=0.5]{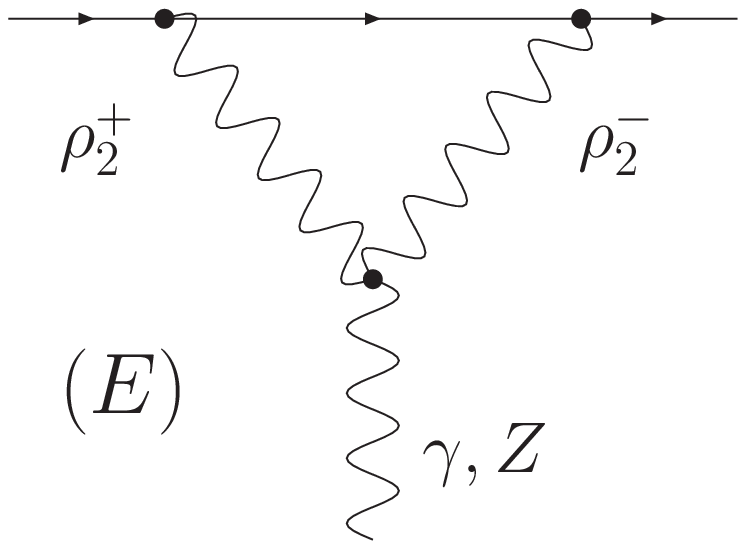}
\includegraphics[bb=179 544 436 727,scale=0.5]{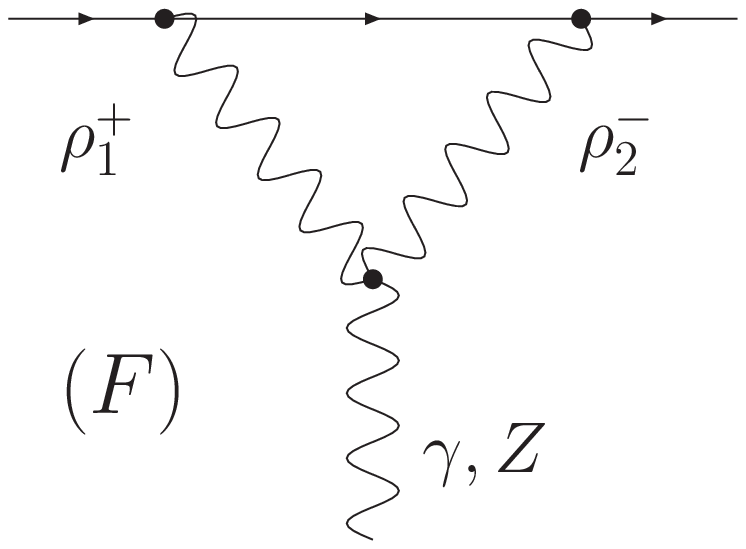} \\ 
\vspace{0.75cm}
\includegraphics[bb=178 538 427 729,scale=0.5]{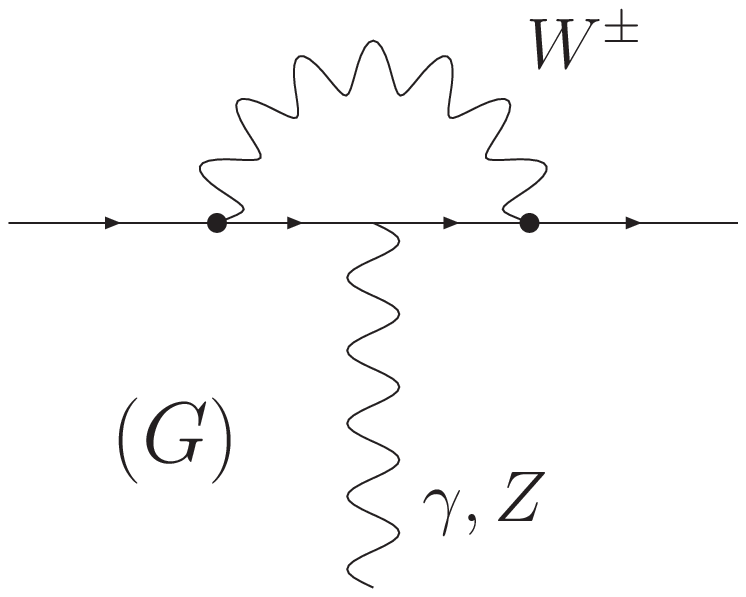}
\includegraphics[bb=178 538 427 729,scale=0.5]{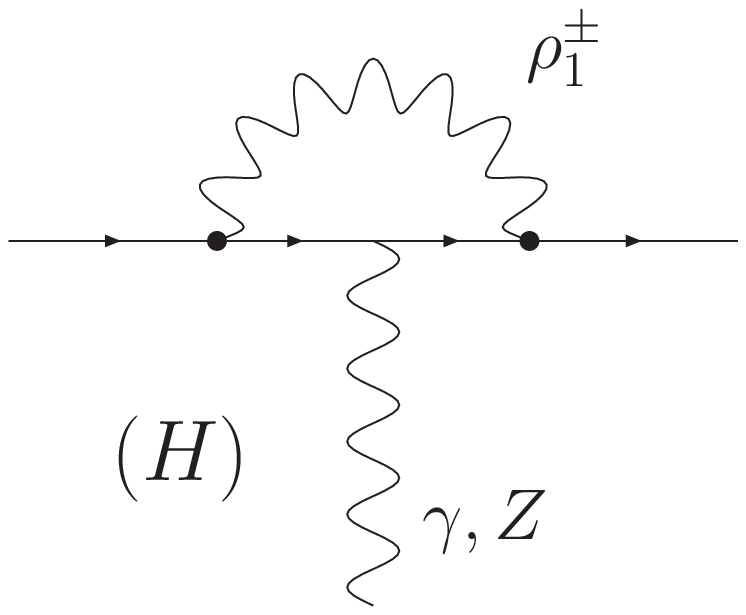}
\includegraphics[bb=178 538 427 729,scale=0.5]{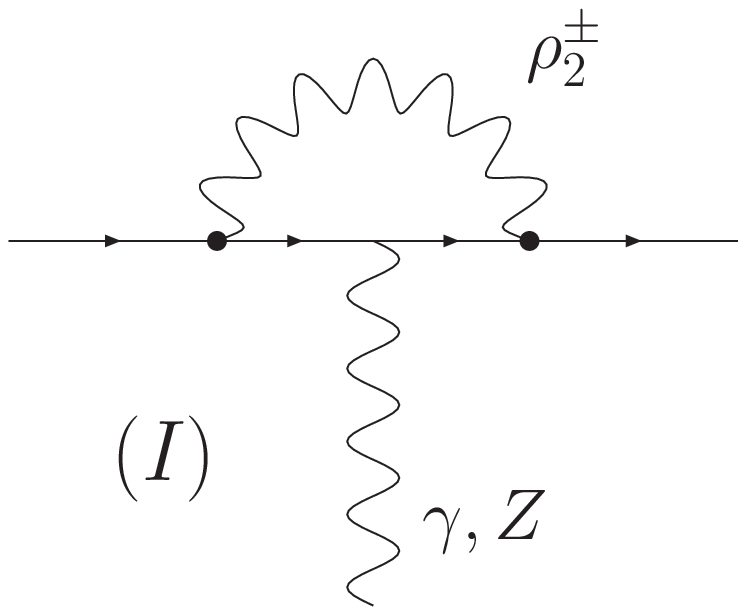} \\ 
\vspace{0.75cm}
\includegraphics[bb=175 532 428 724,scale=0.5]{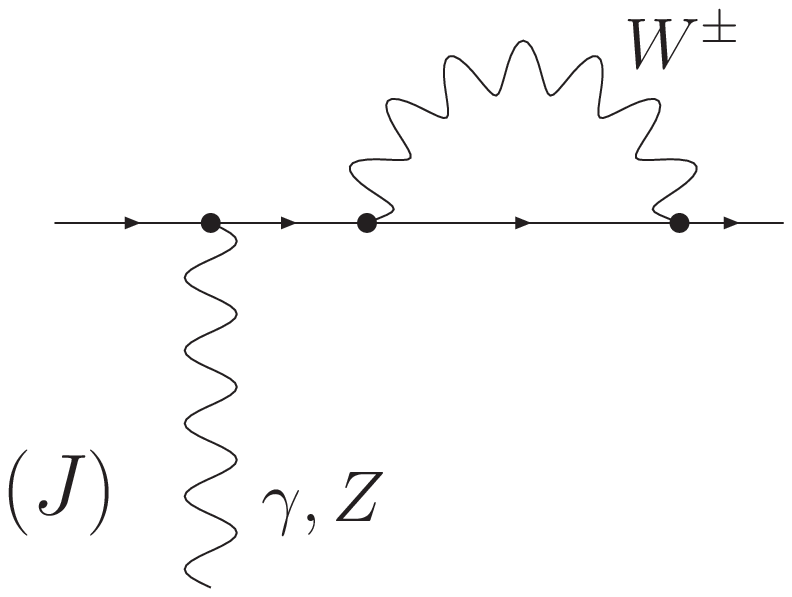}
\includegraphics[bb=175 532 428 724,scale=0.5]{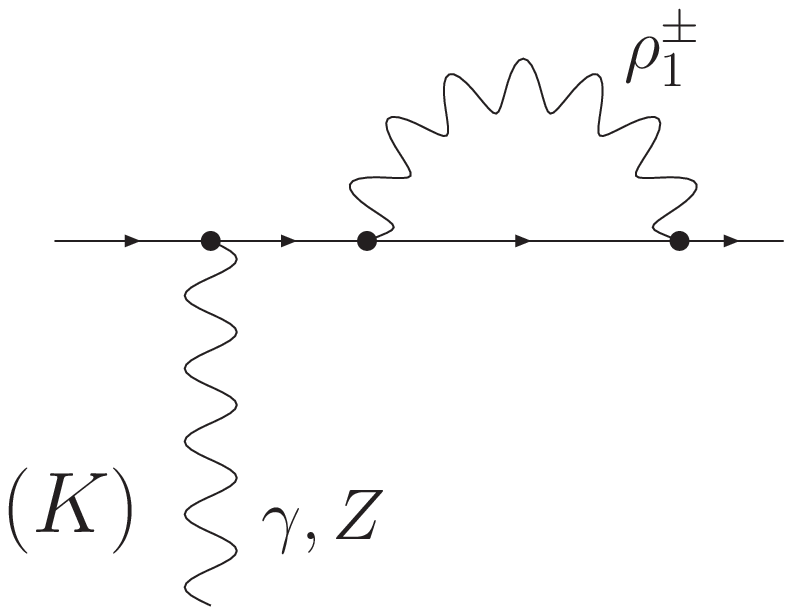}
\includegraphics[bb=175 532 428 724,scale=0.5]{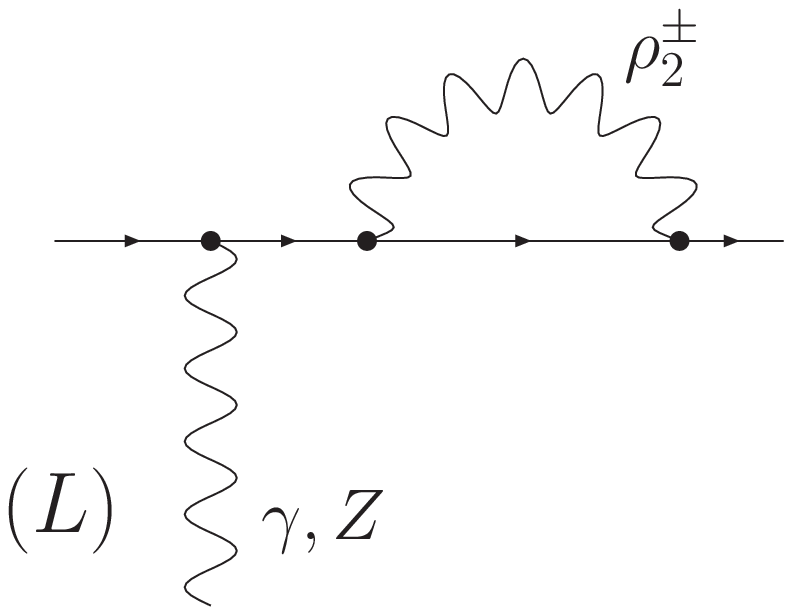}
\end{center}
\caption[]{One-loop vertex corrections which contribute to the gauge-invariant
  self-energies in the four-site model.}
\label{fg:vertex-diagrams}
\end{figure}

The vertex corrections which give rise to pinch contributions are shown in 
Fig.~\ref{fg:vertex-diagrams}. 
We note that contributions
involving internal neutral gauge bosons cancel among themselves (see Eq. (58)
of Ref.~\cite{Dawson:2007yk}).
In deriving the fermion-gauge
boson couplings, we assume that the delocalization parameter,
$x_1$, is of ${\cal O}(x^2)$ (a necessary condition to cancel
the tree level contribution to $S$).
  Using the results of Ref.~\cite{Dawson:2007yk},
we find the chiral logarithmic contributions from the
vertex pinch diagrams are to ${\cal O}(x^0)$:
\begin{eqnarray}
\label{eq:DelPiAA-vertex}
\Delta\Pi_{AA}^{vertex} &=& \frac{\alpha}{4\pi} \biggl[
   \frac{8}{3 c^2} + \frac{1}{6 c^4} \biggr] \log \frac{\Lambda^2}{M_W^2} \,, \\
\nonumber\\
\label{eq:DelPiZA-vertex}
\Delta\Pi_{ZA}^{vertex} &=& \frac{\alpha}{4\pi sc} \biggl[
   c^2 \biggl(\frac{4}{3c^2} + \frac{1}{12c^4}\biggr) \log \frac{\Lambda^2}{M_W^2}
  - \frac{3}{4}(1+z^2)\left(1 - \frac{x_1}{ x^2}\right)
   \log \frac{\Lambda^2}{M_{\rho_1^\pm}^2} \biggr] \,,\\
\nonumber\\
\label{eq:DelPiZZ-vertex-flat-local}
\Delta\Pi_{ZZ}^{vertex} &=& \frac{\alpha}{4\pi s^2 c^2} \left[
  -\frac{3}{2} c^2 (1+z^2)\left(1 - \frac{x_1}{ x^2}\right)
   \right] \log \frac{\Lambda^2}{M_{\rho_1^\pm}^2} \,.  
\end{eqnarray}

\begin{figure}[t]
\begin{center}
\includegraphics[bb=180 522 425 729,scale=0.5]{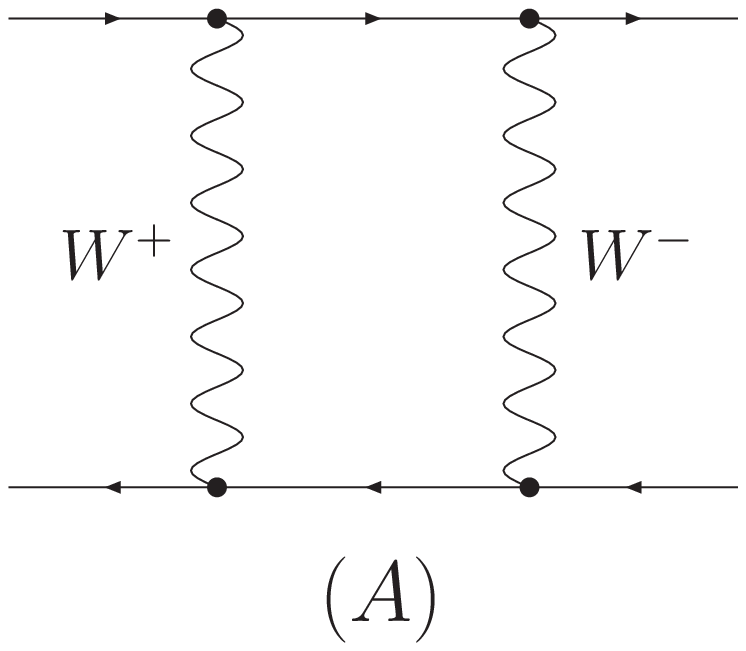} 
\includegraphics[bb=180 522 425 729,scale=0.5]{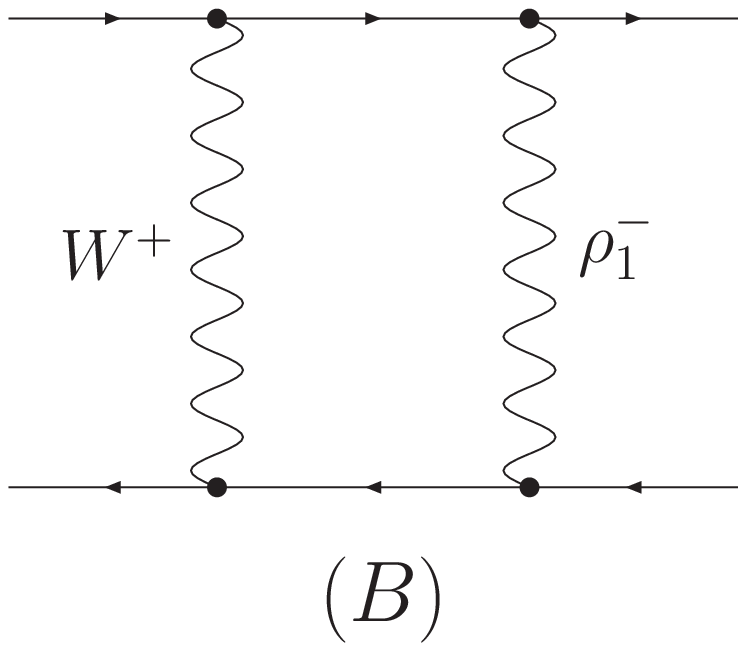}
\includegraphics[bb=180 522 425 729,scale=0.5]{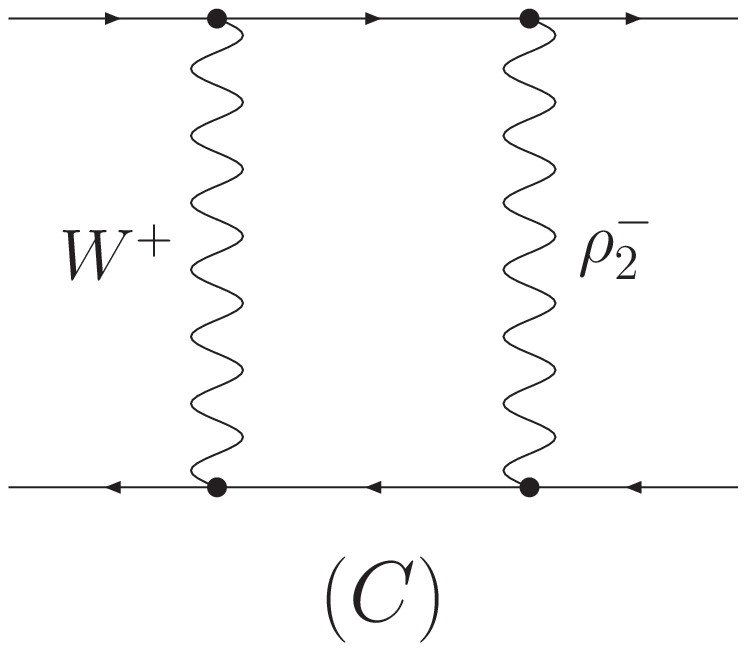} \\ 
\vspace{0.75cm}
\includegraphics[bb=180 522 425 729,scale=0.5]{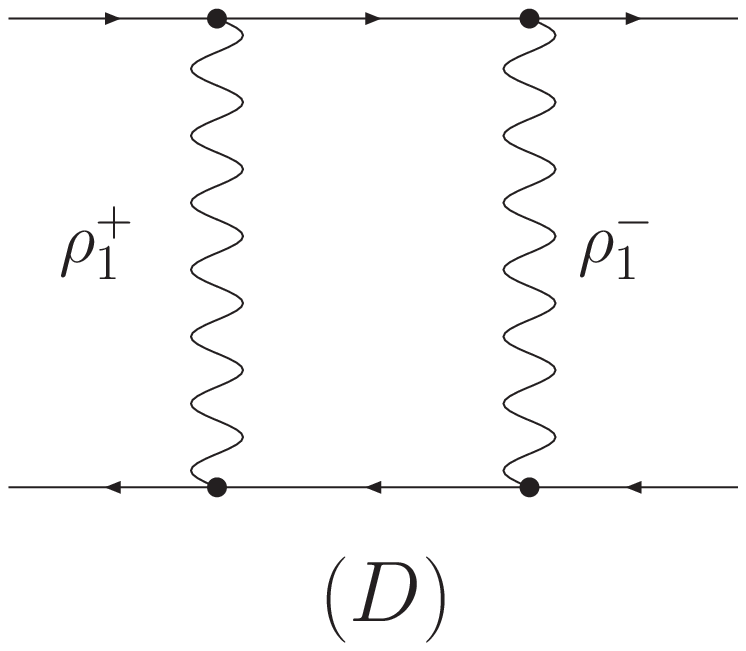}
\includegraphics[bb=180 522 425 729,scale=0.5]{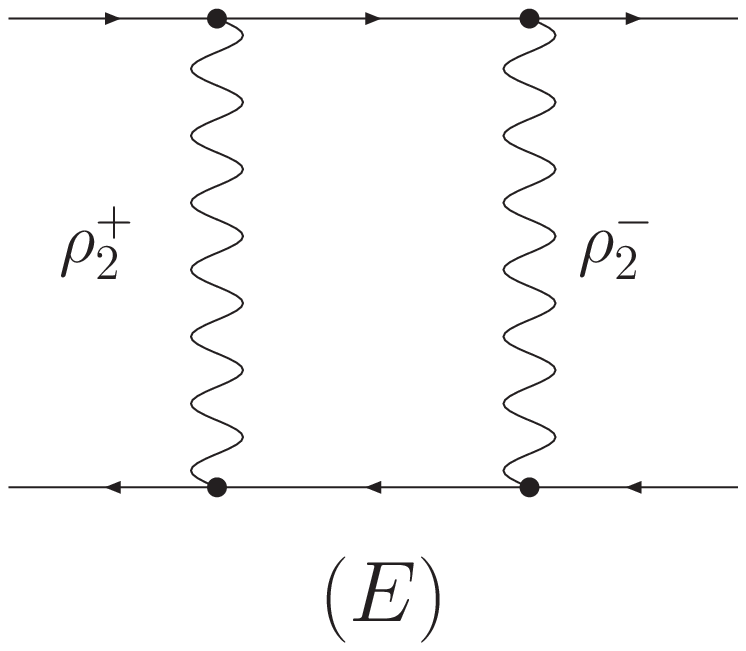}
\includegraphics[bb=180 522 425 729,scale=0.5]{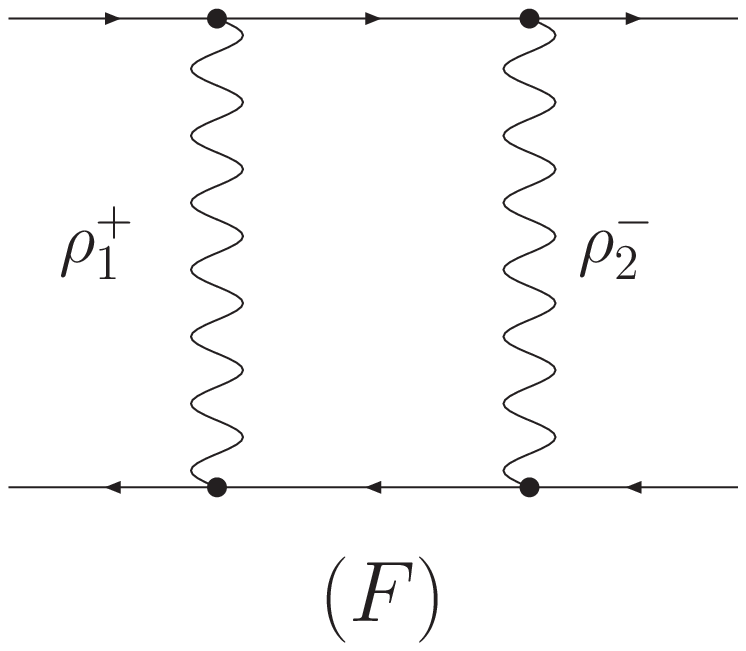}
\end{center}
\caption[]{One-loop box corrections which contribute to the gauge-invariant
  self-energies in the four-site model.}
\label{fg:box-diagrams}
\end{figure}

The box corrections which contribute to the pinch pieces are shown in 
Fig.~\ref{fg:box-diagrams}.  Summing the individual diagrams, we find
the total contributions are:
\begin{eqnarray}
\label{eq:DelPiAA-box}
\Delta\Pi_{AA}^{box} &=& \frac{\alpha}{4\pi} \biggl[
   -\frac{3}{2 c^2} - \frac{1}{12 c^4} \biggr] \log \frac{\Lambda^2}{M_W^2} \,, \\
\nonumber\\
\label{eq:DelPiZA-box}
\Delta\Pi_{ZA}^{box} &=& 0 \,, \\
\nonumber\\
\label{eq:DelPiZZ-box}
\Delta\Pi_{ZZ}^{box} &=& \frac{\alpha}{4\pi s^2 c^2} \biggl(
  \frac{3c^2}{2} \biggr) \log \frac{\Lambda^2}{M_W^2} \,.  
\end{eqnarray}

Finally, summing the two-point, vertex and box contributions, we find the 
gauge-independent PT self-energies are:
\begin{eqnarray}
\label{eq:DelPiAA-PT}
\Delta\Pi_{AA}^{PT} &=& \frac{\alpha}{4\pi} \left[
  7 \log \frac{\Lambda^2}{M_W^2} + 7 \log \frac{\Lambda^2}{M_{\rho_1^\pm}^2} +
  7 \log \frac{\Lambda^2}{M_{\rho_2^\pm}^2} \right] \,,\\
\nonumber\\
\Delta\Pi_{ZA}^{PT} &=& \frac{\alpha}{4\pi sc} \biggl[
  c^2 \biggl(7 + \frac{1}{6c^2}\biggr) \log \frac{\Lambda^2}{M_W^2} \nonumber\\
&& \,\,\,\,\,\,\,
  + \left( \frac{7(c^2 - s^2)}{2} - \frac{3}{4}(1+z^2)\left(1 - \frac{x_1}{ x^2}\right) 
  \right) \log \frac{\Lambda^2}{M_{\rho_1^\pm}^2} \nonumber\\
&& \,\,\,\,\,\,\,
\label{eq:DelPiZA-PT}
  + \frac{7(c^2 - s^2)}{2} \log \frac{\Lambda^2}{M_{\rho_2^\pm}^2} \biggr] \,, \\
\nonumber\\
\Delta\Pi_{ZZ}^{PT} &=& \frac{\alpha}{4\pi s^2 c^2} \biggl[
  c^2 \biggl(7c^2 + \frac{1}{3} - \frac{1}{12c^2}\biggr) \log \frac{\Lambda^2}{M_W^2} \nonumber\\
&& \,\,\,\,\,\,\,\,\,\,\,\,
  + \left( \frac{17}{24}\frac{(1-z^4)^2}{(1-z^2)} + \frac{7(c^2-s^2)^2}{4} -
    \frac{3}{2} c^2 (1+z^2) \left(1 - \frac{x_1}{ x^2} \right)
   \right) \log \frac{\Lambda^2}{M_{\rho_1^\pm}^2} \nonumber\\
&& \,\,\,\,\,\,\,\,\,\,\,\,
\label{eq:DelPiZZ-PT}
  + \left( \frac{17}{24}z^2 (1+z^4) + \frac{25}{12}z^4 + \frac{7(c^2-s^2)^2}{4}
    \right)\log \frac{\Lambda^2}{M_{\rho_2^\pm}^2} \biggr] \,.
\end{eqnarray}
%

\section{The $S$ Parameter at One-loop}
\label{sec:Sparam-oneloop}

In this section, we compute the $S$ parameter at the one-loop level in the 
four-site model.  First, let us consider the contribution at tree-level.
In general, the tree-level contribution to the $S$ parameter from an
$SU(2)^{N+1} \times U(1)$ deconstructed Higgsless model with one-site
fermion delocalization is given by \cite{Chivukula:2005bn}:
\begin{equation}
\alpha S_{tree} = 4 s^2 c^2 M_Z^2 \left( \sum_{i = 1}^N \frac{1}{M_{\rho_i^0}^2} - 
  x_1 \tilde{\bf{m}}^{-2}
  \right) \,,
\end{equation}
where $\tilde{\bf{m}}$ is the (0,0) component of the neutral gauge boson mass
matrix.  In particular, for the four-site model, we have:
\begin{eqnarray}
\alpha S_{tree} &=& \frac{4 s^2 M_W^2}{M_{\rho_1^0}^2} \left[
  1 + \frac{M_{\rho_1^0}^2}{M_{\rho_2^0}^2} - x_1 \left( 
  \frac{4 M_{\rho_1^0}^2}{g^2 f_1^2} \right) \right] \nonumber\\
\nonumber\\
&\simeq& \frac{4 s^2 M_W^2}{M_{\rho_1^\pm}^2}
  \biggl(1 + z^2 - \frac{x_1 M_{\rho_1^\pm}^2}{2 M_W^2}(1-z^2) \biggr) +
  {\cal{O}}(x^2) \,,
\label{eq:Stree-delocal}
\end{eqnarray}
where $x_1$ measures the amount of delocalization of the light fermions.
It is easy to see that one can exactly cancel the large tree-level contribution
to $S$ if:
\begin{equation}
x_1 = \frac{2 M_W^2 (1+z^2)}{M_{\rho_1^\pm}^2 (1-z^2)} \,.
\label{eq:x1-ideal}
\end{equation}
This situation is termed {\it{ideal delocalization}} \cite{Chivukula:2005xm}.

The one-loop corrections to $S$ for delocalized fermions are given by substituting
Eqs.~(\ref{eq:DelPiAA-PT})-(\ref{eq:DelPiZZ-PT}) into 
Eq.~(\ref{eq:Sdef}).  Doing this, we find:
\begin{eqnarray}
\Delta S &=& \frac{1}{12\pi}\log \frac{\Lambda^2}{M_W^2} - 
   \left[\frac{(43 + z^2 + 17z^4 + 17z^6)}{24\pi} - 
   \frac{3}{4\pi}\frac{x_1}{x^2} (1+z^2) \right] \log \frac{\Lambda^2}{M_{\rho_1^\pm}^2} 
\nonumber\\
&& \,\,\,\,\,\,\,\,\,\,\,
 - \left[\frac{(42 - 17z^2 - 50z^4 - 17z^6)}{24\pi} \right]
   \log \frac{\Lambda^2}{M_{\rho_2^\pm}^2} \,,
\end{eqnarray}
which can be written in the more suggestive form:
\begin{eqnarray}
\Delta S &=& \frac{1}{12\pi}\log \frac{M_{\rho_1^\pm}^2}{M_W^2} - 
   \left[\frac{(41 + z^2 + 17z^4 + 17z^6)}{24\pi} - 
   \frac{3}{4\pi}\frac{x_1}{x^2} (1+z^2) \right] 
   \log \frac{M_{\rho_2^\pm}^2}{M_{\rho_1^\pm}^2} 
\nonumber\\
&& \,\,\,\,\,\,\,\,\,\,\,
 - \left[\frac{(83 - 16z^2 - 33z^4)}{24\pi}  - 
   \frac{3}{4\pi}\frac{x_1}{x^2} (1+z^2) \right]
   \log \frac{\Lambda^2}{M_{\rho_2^\pm}^2} \,.
\label{eq:Sloop}
\end{eqnarray}
We note that the first term in Eq.~(\ref{eq:Sloop}), 
which arises from scaling between $M_W$ and 
$M_{\rho_1^\pm}$, has the same coefficient as the leading chiral-logarithmic
contribution from a heavy Higgs boson:
\begin{equation}
S_{Higgs} = \frac{1}{12\pi} \log \frac{M_H^2}{M_W^2} \,.
\label{eq:Shiggs}
\end{equation}
This is expected, however, since the gauge- and chiral-symmetries of the four-site
model in this energy range are the same as the SM with a heavy Higgs boson
\cite{Peskin:1991sw}.  Since experimental limits on $S$ assume the existence 
of a fundamental Higgs boson, in order to compare our prediction with data, 
we must subtract Eq.~(\ref{eq:Shiggs}) from Eq.~(\ref{eq:Sloop}).

An important check on our calculation is provided by considering the limit
$f_2 \to \infty$ where the four-site model reduces to the three-site model
\cite{SekharChivukula:2008gz}.  
In this limit, we see from Eq.~(\ref{eq:zdef}) that $z \to 0$
and from Eq.~(\ref{eq:Mrhopm2warped}) that $M_{\rho_2^\pm} \to \infty$.  In this
situation, we identify $M_{\rho_2^\pm}$ with the cutoff of the effective theory
($\Lambda$) and Eq.~(\ref{eq:Sloop}) reduces to: 
\begin{equation}
\Delta S = \frac{1}{12\pi}\log \frac{M_{\rho_1^\pm}^2}{M_W^2} - 
   \left[\frac{41}{24\pi} - \frac{3}{4\pi}\frac{x_1}{x^2} \right] 
   \log \frac{\Lambda^2}{M_{\rho_1^\pm}^2} \,.
\label{eq:Sloop-3site}
\end{equation}
This expression is exactly the one obtained in Refs.~\cite{Matsuzaki:2006wn,
SekharChivukula:2007ic} for the three-site model (using two different 
methods and two different gauges) and checked numerically in 
Ref.~\cite{Dawson:2007yk}.  We note that, originally, this term accounted
for scaling between $M_{\rho_1^\pm}$ and $M_{\rho_2^\pm}$, i.e., the
energy range where the gauge- and chiral-symmetries of the four-site model 
are identical to those of the three-site model. 

As mentioned in Section~\ref{sec:the-model}, the $S$ parameter also receives contributions
from the dimension-4 operators of Eq.~(\ref{eq:L4doperators}) \cite{Perelstein:2004sc}.  
By using Eqs.~(\ref{eq:L3eigen})-(\ref{eq:Beigen}), these may be written as
\cite{Matsuzaki:2006wn}:
\begin{equation}
{\cal{L}}_4 |^{quad}_{Z,A} = 
  \frac{i}{2} \delta_{ZZ} (Z_\mu D^{\mu\nu} Z_\nu) + 
  i \delta_{ZA} (Z_\mu D^{\mu\nu} A_\nu) +
  \frac{i}{2} \delta_{AA} (A_\mu D^{\mu\nu} A_\nu) \,,
\label{eq:L4masseigen}
\end{equation}
where $D^{\mu\nu} = - \partial^2 g^{\mu\nu} + \partial^\mu \partial^\nu$.  We find 
the $\delta_{ij}$'s in the four-site model take the form:
\begin{eqnarray}
\label{eq:deltaZZ}
\delta_{ZZ} &=& \frac{e^2}{s^2 c^2} \left[c_1 s^2 (z^2 - (c^2-s^2)) +
  c_2 c^2 (z^2 + (c^2-s^2)) - \frac{1}{2}c_3 (z^4 - (c^2-s^2)^2) \right] \,, \\
\nonumber\\
\label{eq:deltaZA}
\delta_{ZA} &=& \frac{e^2}{2sc} \left[c_1 (-z^2 + (c^2-3s^2)) + 
  c_2 (z^2 + (3c^2 - s^2)) + 2 c_3(c^2 - s^2) \right] \, \\
\nonumber\\
\label{eq:deltaAA}
\delta_{AA} &=& 2e^2 \left[c_1 + c_2 + c_3 \right] \,.
\end{eqnarray}
Inserting these expressions into Eq.~(\ref{eq:Sdef}), we find the contribution
to $S$ from the dimension-4 operators is:
\begin{equation}
\delta S_{1-loop} = -8\pi \left[
  (1-z^2)(c_1 + c_2) + (1 + z^4) c_3 \right] \,.
\label{eq:deltaS-CT}
\end{equation}

An interesting question to ask in models with delocalized fermions is whether or not
the {\it{ideal}} value of $x_1$ which cancels the large tree-level contribution to
the $S$ parameter is {\it{ideal}} at higher-orders in perturbation theory.  In other
words, the question is whether or not $x_1$ must be tuned order-by-order in perturbation
theory in such a way to bring $S$ into agreement with precision electroweak data.

\begin{figure}[t]
\begin{center}
\includegraphics[scale=0.5]{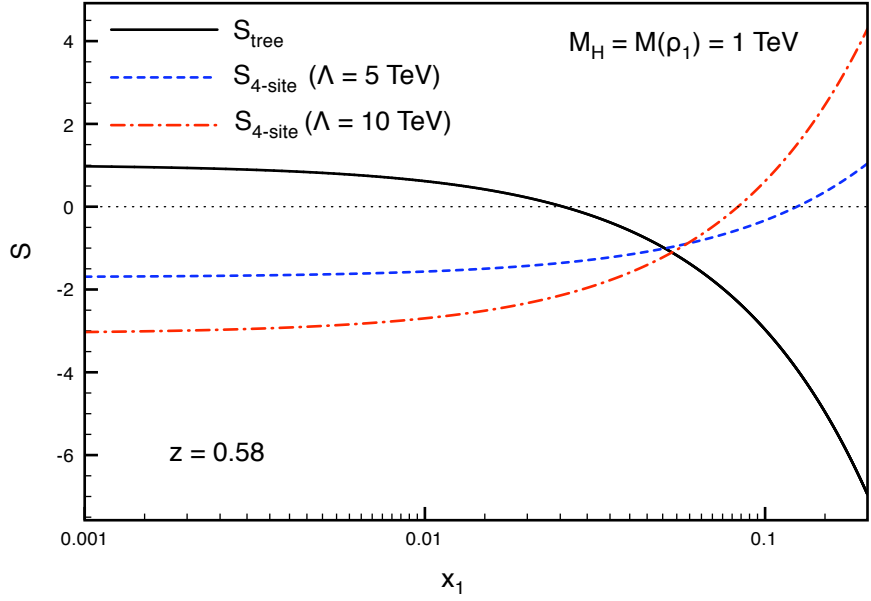}
\end{center}
\caption[]{The $S$ parameter in the four-site model with delocalized fermions at
the one-loop level as a function of the delocalization parameter $x_1$.  In this 
plot, $f_2 = f_1$.}
\label{fg:Sparam4site-x1dep2}
\end{figure}

\begin{figure}[t]
\begin{center}
\includegraphics[scale=0.5]{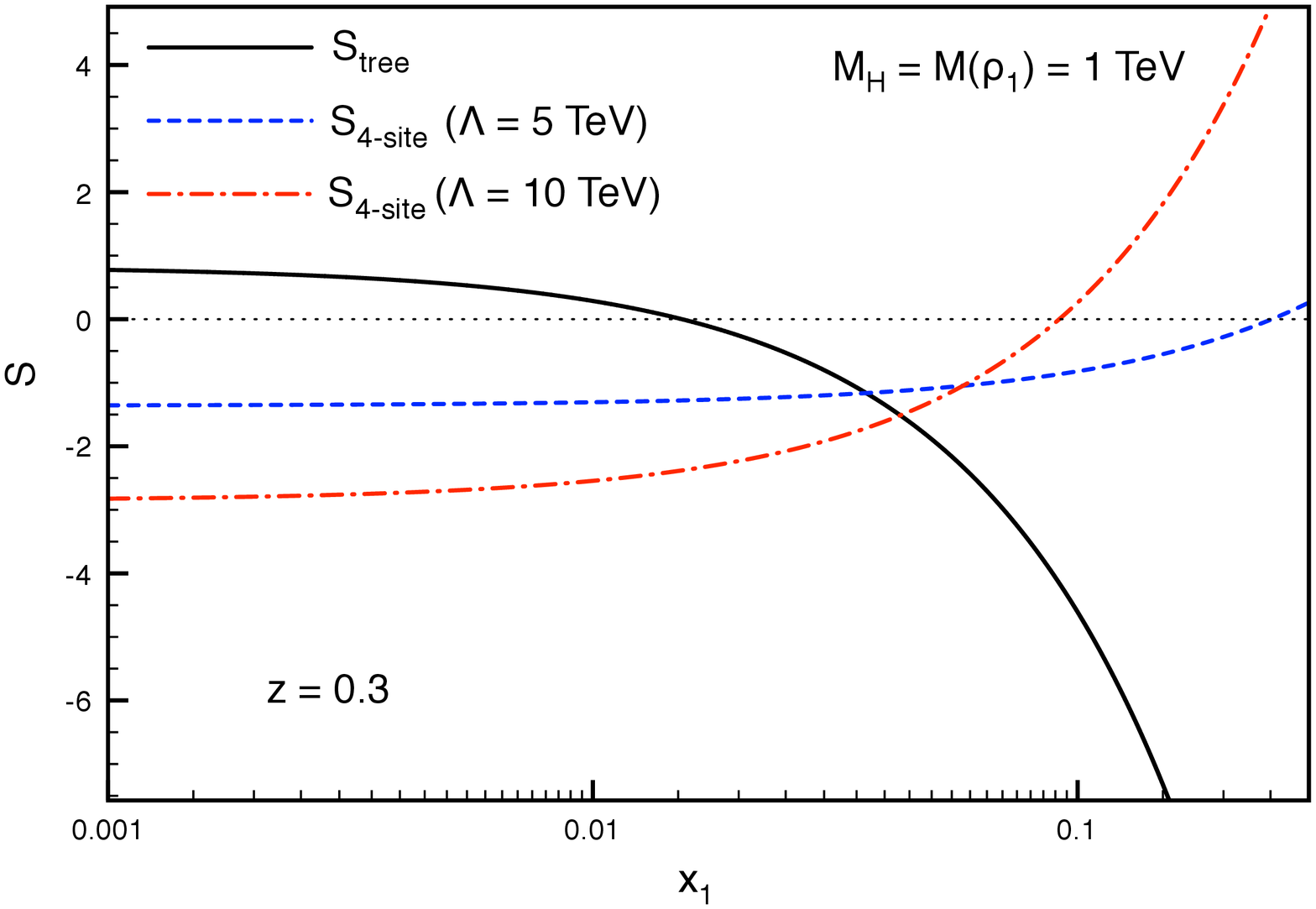}
\end{center}
\caption[]{The $S$ parameter in the four-site model with delocalized fermions at
the one-loop level as a function of the delocalization parameter $x_1$.  In this 
plot, $f_2 \ne f_1$.}
\label{fg:Sparam4site-x1dep}
\end{figure}

To study this issue in the four-site model, we apply the following numerical 
analysis.  The model initially has five input parameters: the three gauge couplings
and the two link parameters, along with the cutoff, $\Lambda$.
 We have taken as our input
parameters $\alpha$, $G_F$ and $M_Z$ and computed the weak mixing angle as
described in Appendix~\ref{app:EW-parameters}.  The two remaining parameters
we choose as $M_{\rho_1^\pm}$ and $z$.
In Fig.~\ref{fg:Sparam4site-x1dep}, we plot the tree-level expression for $S$ 
(Eq.~(\ref{eq:Stree-delocal})) as well as the sum of the tree-level and one-loop 
contributions (with $c_1 = c_2 = 0$) as a function of $x_1$.  In this plot, we 
have taken $z=0.58$ (corresponding to the flat $f_2 = f_1$ scenario) and
we have identified the Higgs reference mass with 
$M_{\rho_1^\pm}$ and assumed two values of the cutoff scale. 
 The horizontal dotted
line in the plot denotes the value $S = 0$.  As we can see, in going from tree-level 
to the one-loop level, the value of $x_1$ must be tuned by factors of $\sim 5$
or more depending on the value of $\Lambda$ and there are potentially large
cancellations between the tree and the one-loop contributions.  
The result for a warped case ($f_1 \ne f_2$) is shown in 
Fig.~\ref{fg:Sparam4site-x1dep2} and looks quite similar to the flat case.

Finally, summing Eqs.~(\ref{eq:Stree-delocal}), (\ref{eq:Sloop}) and 
(\ref{eq:deltaS-CT}), and accounting for the reference Higgs mass, we find our
final result to be:
\begin{eqnarray}
\alpha S_{4-site} &=& \biggl[\frac{4 s^2 M_W^2}{M_{\rho_1^\pm}^2}
  \biggl(1 + z^2 - \frac{x_1 M_{\rho_1^\pm}^2}{2 M_W^2}(1-z^2) \biggr) \biggr]_{\mu=\Lambda} +
  \frac{\alpha}{12\pi} \log \frac{M_{\rho_1^\pm}^2}{M_{H_{ref}}^2} \nonumber\\
&& 
 - 
   \frac{\alpha}{\pi} \left[\frac{(41 + z^2 + 17z^4 + 17z^6)}{24} - 
   \frac{3}{4}\frac{x_1}{x^2} (1+z^2) \right] \log \frac{M_{\rho_2^\pm}^2}{M_{\rho_1^\pm}^2} 
\nonumber\\
&& 
 - \frac{\alpha}{\pi} \left[\frac{(83 - 16z^2 - 33z^4)}{24} - 
   \frac{3}{4}\frac{x_1}{x^2} (1+z^2) \right]
   \log \frac{\Lambda^2}{M_{\rho_2^\pm}^2} \nonumber\\
&& 
 -8\pi \alpha \left[
   (1-z^2)(c_1(\Lambda) + c_2(\Lambda)) + (1 + z^4) c_3(\Lambda) \right] \,,
\label{eq:S4site}
\end{eqnarray}
where the contributions from the tree-level and dimension-four operators are now
understood to be evaluated at the scale $\mu = \Lambda$.

\section{Discussion and Conclusions}

In this paper, we have computed the leading chiral-logarithmic corrections to the
$S$ parameter in the four-site model.  The gauge sector of this model consists
of a SM-like set of of gauge bosons (massless photon and light vector gauge bosons
$W^\pm$ and $Z$) plus two sets of heavier gauge bosons ($\rho_i^\pm$ and 
$\rho_i^0$ with $i$ = 1,2).  Thus, the spectrum is very similar to that of the
lightest and next-to-lightest KK excitations of a Randall-Sundrum scenario 
with an $SU(2)_L \times SU(2)_R \times U(1)_X$ bulk gauge symmetry
\cite{Agashe:2003zs}. 

Our results show that the $S$ parameter in this model is UV-sensitive and 
therefore requires renormalization.  This is similar to the situation in 
Ref.~\cite{Burdman:2008gm} where the one-loop corrections to $S$ from the 
Higgs sector of a holographic model were computed and found to be 
logarithmically-divergent.

We find that the chiral-logarithmic corrections to $S$ in the four-site model
naturally separate into three distinct contributions: a model-independent piece 
arising from scaling from $M_W$ up to the $\rho_1^\pm$ mass, a piece which 
arises from scaling between $M_{\rho_1^\pm}$ and $M_{\rho_2^\pm}$ and a piece
arising from scaling from $M_{\rho_2^\pm}$ up to the cutoff of the
effective theory ($\Lambda$).  The coefficient of the model-independent term 
has the same form as the large Higgs-mass dependence of the $S$ parameter 
in the SM.  This allows us to compare our one-loop results directly with 
experimental constraints on $S$.  Additionally, in the limit $f_2 \to \infty$ where the
four-site model reduces to the three-site model, we have shown that our results
correctly reproduce the one-loop corrections to the $S$ parameter in the 
three-site model \cite{Matsuzaki:2006wn,SekharChivukula:2007ic,Dawson:2007yk}.

In this work, we have focussed on the contributions to $S$ at one-loop primarily
from the gauge bosons of the model.  In principle, there would be contributions
from the extended fermion sector of the model as well.  However, these contributions
have been computed in the three-site model and shown to be negligible 
\cite{Matsuzaki:2006wn}.

We have also studied the dependence of the one-loop results on the delocalization
parameter $x_1$.  In an {\it{ideally-delocalized}} situation, the large 
tree-level contribution to $S$ present in the four-site model can be completely 
cancelled.  The outstanding
issue in these types of models, however, is whether or not $x_1$ must be
tuned order-by-order in perturbation theory to bring $S$ into agreement with
experimental constraints.  We have shown that, in going from tree-level to 
the one-loop level, the {\it{ideal}} value of $x_1$ must be tuned by a factor 
of 5 or more depending on the value of the cutoff scale $\Lambda$.

\section*{Acknowledgements}
The work of S.D. (C.J.) is supported by the U.S. Department of Energy under 
grant DE-AC02-98CH10886 (DE-AC02-06CH11357).

\appendix

\section{Eigenvector Components}
\label{app:eigenvec-components}

The eigenvector components for the charged gauge bosons (defined through Eqs.
(\ref{eq:Lpmeigen})-(\ref{eq:V2pmeigen})) are\footnote{
Our results agree with Ref. \cite{Accomando:2008jh} except for Eq. A7, where
we are a factor of $\sqrt{2}$ larger, and Eq. A8, where our coefficient
for the ${\cal O}(x^2)$ term is a factor of $4$ smaller.}:
\begin{eqnarray}
\label{eq:vwL}
v_{W^\pm}^L &\simeq& 1 - \frac{x^2 z_W}{2} \,,\\
\nonumber\\
\label{eq:vrhopm1L}
v_{\rho_1^\pm}^L &\simeq& -\frac{x}{\sqrt{2}} 
  \biggl(1 + \frac{x^2}{4} \frac{1-3z^2}{1-z^2} \biggr) \,,\\
\nonumber\\
\label{eq:vrhopm2L}
v_{\rho_2^\pm}^L &\simeq& -\frac{x z^2}{\sqrt{2}} 
  \biggl(1 + \frac{x^2 z^2}{4} \frac{3z^4 - 5z^2 + 4}{1-z^2} \biggr) \,, \\
\nonumber\\
\label{eq:vWV1}
v_{W^\pm}^{V_1} &\simeq& \frac{(1+z^2) x}{2} 
  \biggl(1 + \frac{x^2}{4} \frac{(1-3z^2)(1+z^4)}{(1+z^2)} \biggr) \,, \\
\nonumber\\
\label{eq:vrhopm1V1}
v_{\rho_1^\pm}^{V_1} &\simeq& \frac{1}{\sqrt{2}}
  \biggl(1 - \frac{x^2}{4} \frac{1+z^2}{1-z^2} \biggr) \,, \\
\nonumber\\
\label{eq:vrhopm2V1}
v_{\rho_2^\pm}^{V_1} &\simeq& \frac{1}{\sqrt{2}}
  \biggl(1 + \frac{x^2 z^4}{4} \frac{1+z^2}{1-z^2} \biggr) \,, \\
\nonumber\\
\label{eq:vWV2}
v_{W^\pm}^{V_2} &\simeq& \frac{(1-z^2) x}{2} 
  \biggl(1 + \frac{x^2}{4}(1 - 3z^4) \biggr) \,,\\
\nonumber\\
\label{eq:vrhopm1V2}
v_{\rho_1^\pm}^{V_2} &\simeq& \frac{1}{\sqrt{2}}
  \biggl(1 - \frac{x^2}{4} \frac{1-3z^2}{1-z^2} \biggr) \,,\\
\nonumber\\
\label{eq:vrhopm2V2}
v_{\rho_2^\pm}^{V_2} &\simeq& -\frac{1}{\sqrt{2}}
  \biggl(1 - \frac{x^2 z^4}{4} \frac{3-z^2}{1-z^2} \biggr) \,.
\end{eqnarray}

For the neutral gauge bosons, the eigenvector components are\footnote{
Our results agree with Ref. \cite{Accomando:2008jh}.}:
\begin{eqnarray}
v_A^L &\simeq& s (1 - x^2 s^2) \,,\\
v_Z^L &\simeq& c \biggl[ 1 - \frac{x^2}{4c^2} (1 + z^4 - 4s^4) \biggr] \,,\\
v_{\rho^0_1}^L &\simeq& -\frac{x}{\sqrt{2}} \biggl[ 1 - \frac{x^2}{4 c^2}
  \biggl(1 - \frac{2(1-2z^2)\cos 2 \theta}{1-z^2}\biggr) \biggr] \,,\\
v_{\rho^0_2}^L &\simeq& -\frac{xz^2}{\sqrt{2}} \biggl[ 1 + \frac{x^2 z^2}{4 c^2}
  \biggl(2 - 3z^2 + \frac{2\cos 2\theta}{1-z^2} \biggr) \biggr] \,,\\
\nonumber\\
v_A^{V_1} &\simeq& xs(1 - x^2s^2) \,,\\
v_Z^{V_1} &\simeq& \frac{x(z^2 + \cos2\theta)}{2c}\left[
  1 - \frac{x^2}{4c^2}\left(\frac{3z^6 - \cos^3 2\theta - (1+2s^2)z^4 
  - (1-4c^2)z^2}{(z^2 + \cos2\theta)}\right) \right] \,,\\
v_{\rho^0_1}^{V_1} &\simeq& \frac{1}{\sqrt{2}} \biggl[1 - \frac{x^2}{4c^2}
  \biggl(1 + \frac{2 z^2 c^2}{1-z^2} \biggr) \biggr] \,, \\
v_{\rho^0_2}^{V_1} &\simeq& \frac{1}{\sqrt{2}} \biggl[1 - \frac{x^2 z^4}{4c^2}
  \biggl(1 - \frac{2 \cos 2\theta}{1-z^2} \biggr) \biggr] \,,\\
\nonumber\\
v_A^{V_2} &\simeq& xs(1 - x^2s^2) \,,\\
v_Z^{V_2} &\simeq& -\frac{x(z^2 - \cos2\theta)}{2c} \biggl[ 1 - \frac{x^2}{4c^2}
  \biggl(\frac{3z^6 - (1+2c^2)z^4 + \cos^3 2\theta - (1-4s^4)z^2}{z^2-\cos2\theta}
  \biggr) \biggr] \,,\\
v_{\rho^0_1}^{V_2} &\simeq& \frac{1}{\sqrt{2}} \biggl[1 - \frac{x^2}{4c^2}
  \biggl(1 - \frac{2z^2 \cos2\theta}{1-z^2} \biggr) \biggr] \,,\\
v_{\rho^0_2}^{V_2} &\simeq& -\frac{1}{\sqrt{2}} \biggl[1 - \frac{x^2 z^4}{4c^2}
  \biggl(1 + \frac{2\cos2\theta}{1-z^2} \biggr) \biggr] \,,\\
\nonumber\\
v_A^{B} &\simeq& c(1 - x^2 s^2) \,,\\
v_Z^{B} &\simeq& -s \biggl[1 - \frac{x^2}{4c^2} (1 + z^4 - 4c^4) \biggr] \,,\\
v_{\rho^0_1}^{B} &\simeq& -\frac{xt}{\sqrt{2}} \biggl[1 - \frac{x^2}{4c^2}
  \biggl(1 + \frac{2(1-2z^2)\cos2\theta}{1-z^2} \biggr) \biggr] \,,\\
v_{\rho^0_2}^{B} &\simeq& \frac{x z^2 t}{\sqrt{2}} \biggl[ 1 + \frac{x^2 z^2}{4c^2}
  \biggl(2 - 3z^2 - \frac{2\cos2\theta}{1-z^2} \biggr) \biggr] \,.
\end{eqnarray}
%

\section{Triple Gauge Boson Interactions and Couplings}
\label{app:TGVcouplings}

Expanding the non-Abelian interaction terms of Eq.~(\ref{eq:Lgauge}), we find the 
triple gauge boson interactions take the form:
\begin{eqnarray} 
{\cal{L}}_{AAA} &=& i g \biggl[(\partial_\mu L^+_\nu - \partial_\nu L^+_\mu) L^{\mu -}L^{\nu 3} +
   \partial_\mu L^3_\nu L^{\mu +} L^{\nu -} \biggr] \nonumber \\
&&  + i \tilde{g} \sum_i^2
  \biggl[(\partial_\mu V^+_{i,\nu} - \partial_\nu V^+_{i,\mu}) V^{\mu -}_i V^{\nu 3}_i +
   \partial_\mu V^3_{i,\nu} V^{\mu +}_i V^{\nu -}_i \biggr] + \mbox{h.c.}\,.
\label{eq:LAAAgauge}
\end{eqnarray}
Inserting the expansions of the gauge eigenstate fields in terms of the mass eigenstate
fields (Eqs.~(\ref{eq:Lpmeigen})-(\ref{eq:V2pmeigen}) and (\ref{eq:L3eigen})-(\ref{eq:V23eigen}))
into this expression yields:
\begin{eqnarray}
{\cal{L}}_{AAA} &=& i \biggl\{
  \sum_{n = A,Z,\rho_1^0,\rho_2^0} g_{W^+ W^-}^n \biggl(W_{\mu\nu}^+ W^{\mu -} n^\nu +
  \frac{1}{2} n_{\mu\nu}W^{+ \mu} W^{- \nu} \biggr) \nonumber\\
&& + g_{W^+ \rho_1^-}^n \biggl( (W^+_{\mu\nu}\rho_1^{\mu -} + \rho_{1,\mu\nu}^+ W^{\mu -})n^\nu +
  \frac{1}{2} n_{\mu\nu} (W^{\mu +} \rho_1^{\nu -} + \rho_1^{\mu +} W^{\nu -} ) \biggr) \nonumber\\
&& + g_{W^+ \rho_2^-}^n \biggl( (W^+_{\mu\nu}\rho_2^{\mu -} + \rho_{2,\mu\nu}^+ W^{\mu -})n^\nu +
  \frac{1}{2} n_{\mu\nu} (W^{\mu +} \rho_2^{\nu -} + \rho_2^{\mu +} W^{\nu -} ) \biggr) \nonumber\\
&& + g_{\rho_1^+ \rho_2^-}^n \biggl( (\rho^+_{1,\mu\nu}\rho_2^{\mu -} + 
  \rho_{2,\mu\nu}^+ \rho_1^{\mu -}) n^\nu +
  \frac{1}{2} n_{\mu\nu} (\rho_1^{\mu +} \rho_2^{\nu -} + \rho_2^{\mu +} \rho_1^{\nu -} ) 
  \biggr) \nonumber\\
&& + g_{\rho_1^+ \rho_1^-}^n \biggl(\rho_{1,\mu\nu}^+ \rho_1^{\mu -} n^\nu +
  \frac{1}{2} n_{\mu\nu}\rho_1^{+ \mu} \rho_1^{- \nu} \biggr) \nonumber\\
&& + g_{\rho_2^+ \rho_2^-}^n \biggl(\rho_{2,\mu\nu}^+ \rho_2^{\mu -} n^\nu +
  \frac{1}{2} n_{\mu\nu}\rho_2^{+ \mu} \rho_2^{- \nu} \biggr)
                   \biggr\} + \mbox{h.c.}\,,
\end{eqnarray}
where $n_{\mu\nu} = \partial_\mu n_\nu - \partial_\nu n_\mu$ and  $n = A, Z, \rho_i, \rho_2$.
Using Eq.~(\ref{eq:e-def}), the couplings between three mass-eigenstate fields are 
expressed by using the wavefunctions from Appendix~\ref{app:eigenvec-components} as
(to ${\cal{O}}(x^2)$):
\begin{eqnarray}
g_{W^+ W^-}^n &=& \frac{e}{s} (1 + s^2 x^2) \biggl[(v_W^L)^2 v_n^L + \frac{1}{x}
  \biggl((v_W^{V_1})^2 v_n^{V_1} + (v_W^{V_2})^2 v_n^{V_2}\biggr) \biggr] \,\\
\nonumber\\
g_{W^+ \rho_1^-}^n &=& \frac{e}{s} (1 + s^2 x^2) \biggl[v_W^L v_{\rho_1^\pm}^L v_n^L + 
  \frac{1}{x} \biggl(v_W^{V_1} v_{\rho_1^\pm}^{V_1} v_n^{V_1} + 
   v_W^{V_2} v_{\rho_1^\pm}^{V_2} v_n^{V_2} \biggr)
  \biggr] \,\\
\nonumber\\
g_{W^+ \rho_2^-}^n &=& \frac{e}{s} (1 + s^2 x^2) \biggl[v_W^L v_{\rho_2^\pm}^L v_n^L + 
  \frac{1}{x} \biggl(v_W^{V_1} v_{\rho_2^\pm}^{V_1} v_n^{V_1} + 
   v_W^{V_2} v_{\rho_2^\pm}^{V_2} v_n^{V_2} \biggr)
  \biggr] \,\\
\nonumber\\
g_{\rho_1^+ \rho_2^-}^n &=& \frac{e}{s} (1 + s^2 x^2) 
  \biggl[v_{\rho_1^\pm}^L v_{\rho_2^\pm}^L v_n^L + 
  \frac{1}{x} \biggl(v_{\rho_1^\pm}^{V_1} v_{\rho_2^\pm}^{V_1} v_n^{V_1} + 
   v_{\rho_1^\pm}^{V_2} v_{\rho_2^\pm}^{V_2} v_n^{V_2} \biggr)
  \biggr] \,\\
\nonumber\\
g_{\rho_1^+ \rho_1^-}^n &=& \frac{e}{s} (1 + s^2 x^2) 
  \biggl[(v_{\rho_1^\pm}^L)^2 v_n^L + \frac{1}{x}
  \biggl((v_{\rho_1^\pm}^{V_1})^2 v_n^{V_1} + 
  (v_{\rho_1^\pm}^{V_2})^2 v_n^{V_2}\biggr) \biggr] \,\\
\nonumber\\
g_{\rho_2^+ \rho_2^-}^n &=& \frac{e}{s} (1 + s^2 x^2) 
  \biggl[(v_{\rho_2^\pm}^L)^2 v_n^L + \frac{1}{x}
  \biggl((v_{\rho_2^\pm}^{V_1})^2 v_n^{V_1} + 
  (v_{\rho_2^\pm}^{V_2})^2 v_n^{V_2}\biggr) \biggr] \,.
\end{eqnarray}
The couplings between three mass eigenstate gauge bosons are summarized in
Table~\ref{tbl:3-point-gauge-couplings}.

\begin{table}
\begin{center} 
\begin{tabular}{|c||c|c|} 
\hline 
$$  & $n=A$ & $n=Z$  \\ 
  \hline\hline  
$g^n_{W^+W^-}$  & $e$ & $\frac{ec}{s}$ \\ 
\hline 
$g^n_{\rho_1^+ \rho_1^-}$  & $e$  & $\frac{e(c^2-s^2)}{2cs}$ \\ 
\hline 
$g^n_{\rho_2^+ \rho_2^-}$  & $e$  & $\frac{e(c^2-s^2)}{2cs}$\\
\hline 
$g^n_{W^+ \rho_1^-}$ & 0 & $-\frac{e x}{s c} \left(\frac{\sqrt{2}(1-z^4)}{4}\right)$ \\
\hline
$g^n_{W^+ \rho_2^-}$ & 0 & 0 \\
\hline
$g^n_{\rho_1^+ \rho_2^-}$ & 0 & $\frac{e z^2}{2 s c}$ \\
\hline
\end{tabular}
\end{center} 
\caption{ The three-point couplings among the gauge fields in the 
  four-site model to lowest-order in $x$.  We have computed, 
  but do not display, the corrections up to ${\cal{O}}(x^3)$. }
\label{tbl:3-point-gauge-couplings}
\end{table}

\section{Feynman Graph Results}
\label{app:feyngraph-results}

In this appendix, we present the chiral
logarithmic contribution to ${\cal O}(x^0)$
from each diagram in 
Fig.~\ref{fg:twopt-diagrams} which contributes to the two-point functions
$\Delta\Pi^{two-pt.}_{AA,ZA,ZZ}$.  The SM contributions in unitary gauge can
be found in the appendix of Ref. \cite{Dawson:1994fa}.

\subsection{Photon Two-Point Amplitude $\Delta\Pi^{two-pt.}_{AA}$}
\label{subsec:PhotonTwoPtAmps}

The non-zero amplitudes which contribute to $\Delta\Pi_{AA}$ are:
\begin{eqnarray}
(\Delta\Pi_{AA})_A &=& \frac{\alpha}{4\pi} \biggl[7 - \frac{7}{6 c^2}
  - \frac{1}{12 c^4} \biggr] \log \frac{\Lambda^2}{M_W^2} \,\\
\nonumber\\
(\Delta\Pi_{AA})_D &=& \frac{\alpha}{4\pi} (7) \log 
  \frac{\Lambda^2}{M_{\rho_1^\pm}^2} \,\\
\nonumber\\
(\Delta\Pi_{AA})_E &=& \frac{\alpha}{4\pi} (7) \log 
  \frac{\Lambda^2}{M_{\rho_2^\pm}^2} \, ,
\end{eqnarray}
where we have neglected terms of ${\cal{O}}(x^2)$.
 
\subsection{$Z$-Photon Mixing Amplitude $\Delta\Pi^{two-pt.}_{ZA}$}
\label{subsec:ZPhotonMixAmps}

The non-zero amplitudes which contribute to $\Delta\Pi^{two-pt.}_{ZA}$ are:
\begin{eqnarray}
(\Delta\Pi_{ZA})_A &=& \frac{\alpha}{4\pi c s}\biggl[ c^2 \biggl(
  7 - \frac{7}{6 c^2}
  - \frac{1}{12 c^4} \biggr) \biggr] \log \frac{\Lambda^2}{M_W^2} \,\\
\nonumber\\
(\Delta\Pi_{ZA})_D &=& \frac{\alpha}{4\pi c s} 
  \biggl( \frac{7 (c^2 - s^2)}{2} \biggr)
  \log\frac{\Lambda^2}{M_{\rho_1^\pm}^2} \,\\
\nonumber\\
(\Delta\Pi_{ZA})_E &=& \frac{\alpha}{4\pi c s} 
  \biggl( \frac{7 (c^2 - s^2)}{2} \biggr)
  \log\frac{\Lambda^2}{M_{\rho_2^\pm}^2} \, .
\end{eqnarray}

\subsection{$Z$ Boson Two-Point Amplitude $\Delta\Pi^{two-pt.}_{ZZ}$}
\label{subsec:ZTwoPtAmps}

The non-zero contributions to $\Delta\Pi^{two-pt.}_{ZZ}$ are:
\begin{eqnarray}
(\Delta\Pi_{ZZ})_A &=& \frac{\alpha}{4\pi s^2 c^2} \biggl[ c^4
  \biggl(7 - \frac{7}{6 c^2}
  - \frac{1}{12 c^4} \biggr) \biggr] \log \frac{\Lambda^2}{M_W^2} \,\\
\nonumber\\
(\Delta\Pi_{ZZ})_B &=& \frac{\alpha}{4\pi s^2 c^2} \left[
  \frac{17}{24} \frac{(1-z^4)^2}{(1-z^2)}
  \right] \log\frac{\Lambda^2}{M_{\rho_1^\pm}^2} \,\\
\nonumber\\
(\Delta\Pi_{ZZ})_D &=& \frac{\alpha}{4\pi s^2 c^2}
  \biggl( \frac{7 (c^2 - s^2)^2}{4} \biggr)
  \log\frac{\Lambda^2}{M_{\rho_1^\pm}^2} \,\\
\nonumber\\
(\Delta\Pi_{ZZ})_E &=& \frac{\alpha}{4\pi s^2 c^2}
  \biggl( \frac{7 (c^2 - s^2)^2}{4} \biggr)
  \log\frac{\Lambda^2}{M_{\rho_2^\pm}^2} \,\\
\nonumber\\
(\Delta\Pi_{ZZ})_F &=& \frac{\alpha}{4\pi s^2 c^2} \left[
  \frac{17}{24} z^2 (1+z^4) + \frac{25}{12} z^4
  \right] \log\frac{\Lambda^2}{M_{\rho_2^\pm}^2} \,.
\end{eqnarray}

\section{Electroweak Parameters in $4$-Site Model}
\label{app:EW-parameters}

In this appendix, we define the relations between the SM input parameters used
in our numerical analysis and the four-site model parameters.  First, we
define the $SU(2)$ and $U(1)$ couplings as $g$ and $g^\prime$ with 
$\frac{s}{c}=\frac{g^\prime}{g}$.  Next, we take as inputs $e$, $M_Z$, $G_\mu$:
\begin{eqnarray}
\alpha &=& \frac{e^2}{4\pi} = 137.035999679^{-1} \nonumber \\
M_Z&=& 91.1875~  \,\,\mbox{GeV} \nonumber \\
G_F&=& 1.166637\times 10^{-5} \,\,\mbox{GeV}^{-2}\,.
\end{eqnarray}

We expand the definition of the electromagnetic coupling $e$ in the 
four-site model in powers of $x$ as:
\begin{eqnarray}
\label{eq:e-def}
\frac{1}{e^2} &\equiv& \frac{1}{g^2} + \frac{2}{\tilde{g}^2} + 
  \frac{1}{g^{\prime 2}}
\nonumber \\
&=& \frac{1}{g^2 s^2} \biggr( 1+2 s^2 x^2\biggr)\,.
\end{eqnarray}
The relationship between the $W^\pm$ and $Z$ boson masses is given by:
\begin{equation}
M_W^2=c^2 M_Z^2 \biggl(1+x^2 (z_Z-z_W)\biggr)
 \end{equation}

Next, in order to derive a relation for $G_F$, we consider the coupling between
the SM-like $W^\pm$ and light fermions.  Using, Eq.~(\ref{eq:Lfermion}), we find:
\begin{eqnarray}
g_L^W &=& g \left[ v_W^L (1-x_1) + \frac{x_1}{x} v_W^{V_1} \right] \nonumber\\
&=& g \left[ 1 - x^2 \left(\frac{3}{4} - \frac{z^4}{4} \right) \right] \,,
\label{eq:gLW}
\end{eqnarray}
where, for simplicity, we have assumed ideal delocalization for the light
fermions.  Then, Eq.~(\ref{eq:gLW}) corresponds to a value for $G_F$ of:
\begin{equation}
\label{eq:GF}
\sqrt{2} G_F = \frac{(g_L^W)^2}{4 M_W^2} = 
  \frac{g^2}{4 M_W^2} \left[ 1 - x^2 \left( \frac{3}{2} - \frac{z^4}{2} \right) \right] \,.
\end{equation}

Finally, we can calculate the ``$Z$ standard'' weak mixing angle $\theta_Z$:
\begin{eqnarray}
s_Z^2 c_Z^2 &=& \frac{e^2}{4\sqrt{2} G_F M_Z^2} \nonumber \\
&=& s^2 c^2 + 2 x^2 s^2 (c^2 - s^2) \left( c^2 - \frac{1}{4}(1+z^4) \right) \,,
\end{eqnarray}
where $s_Z (c_Z)$ = $\sin\theta_Z (\cos\theta_Z)$.  The relationship between the
weak mixing angle $\theta_Z$ and the angle defined in Eq.~(\ref{eq:SMcoups}) is 
then expressed as:
\begin{equation}
s_Z^2 = s^2 + \Delta \,\,\, ; \,\,\, c_Z^2 = c^2 - \Delta
\end{equation}
where:
\begin{equation}
\Delta \equiv 2 x^2 s^2 \left( c^2 - \frac{1}{4}(1+z^4) \right) \,.
\end{equation}
Therefore, we see that the difference between $s^2$ and $s_Z^2$ is of 
${\cal{O}}(x^2)$.

\bibliography{FourSiteModel}

\end{document}